\documentclass[a4paper, 15pt, oneside]{Thesis}  

\graphicspath{{Figures/}}  

\usepackage[square, numbers, comma, sort&compress]{natbib}  
\usepackage{slashed}
\usepackage{amssymb}
\usepackage{amsmath}
\usepackage{hyperref}
\usepackage{setspace}
\usepackage{graphicx}
\usepackage{verbatim}  
\usepackage{vector}  
\hypersetup{urlcolor=blue, colorlinks=true}  

\begin{document}
\frontmatter	  

\title  {Moduli Space of $SU(2)$ Singular Monopole}
\large

\authors  {\texorpdfstring
            {\href{sshah@maths.tcd.ie}{Sarang Shah}}
            {Sarang Shah}
            }
\addresses  {\groupname\\\deptname\\\univname}  
\date       {September 2010}
\subject    {}

\maketitle

\setstretch{1.3}  

\fancyhead{}  
\rhead{\thepage}  
\lhead{}  

\pagestyle{fancy}  

\Declaration{

\addtocontents{toc}{\vspace{1em}}  

I, Sarang Shah, declare that this thesis titled, `Moduli Space of $SU(2)$ Singular Monopole' and the work presented in it are my own. I confirm that:

\begin{itemize} 
\item[\tiny{$\blacksquare$}] This work was done wholly or mainly while in candidature for a research degree at this University.
 
\item[\tiny{$\blacksquare$}] Where any part of this thesis has previously been submitted for a degree or any other qualification at this University or any other institution, this has been clearly stated.
 
\item[\tiny{$\blacksquare$}] Where I have consulted the published work of others, this is always clearly attributed.
 
\item[\tiny{$\blacksquare$}] Where I have quoted from the work of others, the source is always given. With the exception of such quotations, this thesis is entirely my own work.
 
\item[\tiny{$\blacksquare$}] I have acknowledged all main sources of help.
 
\item[\tiny{$\blacksquare$}] Where the thesis is based on work done by myself jointly with others, I have made clear exactly what was done by others and what I have contributed myself.

\item[\tiny{$\blacksquare$}] I, the undersigned, agree that Trinity College Library may lend or copy this thesis upon request.
\\
\end{itemize}

Signed:\\
\rule[1em]{25em}{0.5pt}  
 
Date:7 September 2010 \\
\rule[1em]{25em}{0.5pt}  
}
\clearpage  

\setstretch{2.0}  
\doublespacing
\addtotoc{Summary}  
\summary{
\addtocontents{toc}{\vspace{1em}}  
The $SU(2)$ monopole with a Dirac singularity was constructed in \cite{Durcan2007} and \cite{CherkisDurcan2008}. We study its moduli space by identifying the tangent direction to the moduli space. The tangent vectors to the moduli space are composed of the monopole's phase and translational zero modes. We construct the phase and translational zero modes using the Nahm transform.
 
These zero modes are then used to construct the metric components $g_{00}$ and $g_{0q}$ of the $SU(2)$ singular monopole moduli space, where $0$ denotes the gauge coordinate and $q$ denotes the translational coordinates. We find the moduli space to be the Taub-NUT space.

It has been shown by Nakajima \cite{Nakajima1990} and Maciocia \cite{Maciocia} that there is an isomorphism between the monopole moduli space of regular monopoles and the moduli space of the Nahm data used to construct them. We compute the moduli space of the Nahm data of a singular monopole using the hyperk\"{a}hler quotient construction as described in  \cite{GibbonsEtAl}.  We find it to be isometric to the moduli space of the singular monopole.

Finally we make an explicit connection between the moduli space of the Nahm data and the monopole moduli space using Corrigan's inner product formula \cite{Osborn1981} independently proving this isomorphism.

}
\clearpage

\comtitle{
\addtocontents{toc}{\vspace{1em}}  

}
\clearpage
\acknowledgements{
\addtocontents{toc}{\vspace{1em}}  

I would first like to thank the generous support made possible by the George Mitchell Scholarship of the US-Ireland Alliance for allowing me to pursue this research.

I would like to thank Dr. Sergey Cherkis for his extensive guidance and for his enouragement, motivation, and enthusiasm.

I would also like to thank the staff of the School of Mathematics for their assistance and support as well as the supportive faculty.

Finally, I thank my family and friends for their support during the course of my research. I would especially like to thank Erica Camp for her encouragement and support throughout the year.

}
\clearpage  


\pagestyle{fancy}  

\lhead{\emph{Contents}}  
\tableofcontents  


\setstretch{1.3}  

\addtocontents{toc}{\vspace{2em}}  

\mainmatter	  
\pagestyle{fancy}  

\large
\doublespacing

\chapter{Introduction} 
\label{Chapter1}
\lhead{Chapter 1. \emph{Introduction}} 

\section{Non-Abelian Gauge Theory}

Yang-Mills theory \cite{YangMills1954} is a generalization of Maxwell theory to non-Abelian gauge groups. It is an example of more general gauge theories. In particular we are interested in Yang-Mills-Higgs theory where in addition to the gauge fields we have an adjoint scalar Higgs field.

The Lagrangian density for Yang-Mills-Higgs theory in Euclidean space with gauge group $\mathcal{G}$ is
\begin{equation}
\mathcal{L}= -\frac{1}{2g^2} Tr(F_{\mu\nu}F^{\mu\nu})+\frac{1}{g^2}Tr(D_\mu\Phi D^\mu \Phi)-V(\Phi),
\end{equation}
where $D_\mu \Phi=\partial_\mu \Phi-i[A_\mu,\Phi]$ is the adjoint covariant derivative, $\mu,\nu=0,1,2,3$, and $g$ is the dimensionless gauge coupling constant working in natural units. We adopt the summation notation where repeated indices indicate summation over those indices. The Lagrangian density has dimension of $1/l^4$ where $l$ is the unit of length .
$F$ is the Lie algebra of $\mathcal{G}$ valued field strength tensor of the gauge field defined as $F_{\mu\nu}=\partial_\mu A_\nu-\partial_\nu A_\mu-i[A_\mu,A_\nu]$, where our gauge field $A_\mu$ transforms as a vector under Lorentz transformation and transforms under the gauge group as a connection $A_\mu\to g^{-1}A_\mu g+g^{-1}\partial_\mu g$ where $g$ is a $\mathcal{G}$ valued function for every $x\in\mathbb{R}^3$. $F$ has the dimension of $1/l^2$ and both $A$ and $\Phi$ have dimension of $1/l$. In the language of fibre-bundles we can think of $A$ as a connection on a vector bundle of structure group $\mathcal{G}$ and where our gauge choice for $A$ is the local bundle trivialization and $g$ is a change of trivialization. 

We are in particular interested in static connections in non-abelian gauge theory that are solitonic. These are solutions which do not decay, are massive, and have certain topological properties, such as a topological charge, which ensure that the solution is stable. Solitons which are time independent, localized in $\mathbb{R}^3$(i.e. are particle-like), and have magnetic charge are called monopoles. Monopoles were first considered by Dirac in the context of electromagnetism in 1931 \cite{Dirac1931}. The experimental search for the magnetic monopole has not yielded a conclusive result and the subject remained dormant until monopole-like solutions to gauge theory were found by 't Hooft \cite{tHooft1974} and Polyakov \cite{Polyakov1974} in 1974. Since then the study of the theory of monopoles has continued apace and has yielded further uses and insights into supersymmetric gauge theory and string theory \cite{HananyWitten}. One can find a pedagogical review of monopoles in \cite{Shnir2005}, \cite{TongNotes}, \cite{WeinbergYi}, and \cite{Prasad1980}.

We note the physical importance of the magnitude of the dimensionless coupling constant $g$. The coupling constant determines the strength of the interaction and consequently what tools we can use to investigate the theory. If $g$ is very small then the theory is weakly coupled and could be described using perturbation theory. If $g$ is large then the theory is strongly coupled and non-perturbative techniques would have to be used to describe the theory. The leading semi-classical contribution in these strongly-coupled theories comes from monopoles and instantons. The study of monopoles is therefore important for both weakly and strongly coupled gauge theories.

To compute explicit monopole solutions we will take the BPS limit where $V(\Phi)$ vanishes.

 We also introduce the following notation for this section
\begin{equation}
\left<A,B\right> = \int d^3x Tr(A^\dagger B),
\end{equation}
so in this notation the Lagrangian is 
\begin{equation}
L=\int d^3x \mathcal{L} = -\frac{1}{2g^2}\left<F_{\mu\nu},F^{\mu\nu}\right>+\frac{1}{g^2}\left<D_\mu \Phi, D^\mu \Phi\right>-\int d^3x V(\Phi)
\end{equation}
and has dimension of $1/l$.

Through the calculus of variations we can derive energy functional of the theory to be
\begin{equation}
E=\frac{1}{g^2}\left(\left<B_i,B_i\right>+\left<E_i,E_i\right>+\left<D_i \Phi,D_i \Phi\right>+\left<D_0 \Phi,D_0 \Phi\right>\right),
\label{eqn:EnergyFN}
\end{equation}
where $B_i=\epsilon_{ijk}F_{jk}$, $E_i=F_{0i}$, $i=1,2,3$, and $E$ has dimension of $1/l$. $B_i$ and $E_i$ are both of dimension $1/l^2$.

We can separate the energy functional into kinetic energy and potential energy terms $E=T+U,$ where 
\begin{eqnarray}
T&=&\frac{1}{g^2}\left(\left<E_i,E_i\right>+\left<D_0 \Phi,D_0 \Phi\right>\right),\\
U&=&\frac{1}{g^2}\left(\left<B_i,B_i\right>+\left<D_i \Phi,D_i \Phi\right>\right).
\end{eqnarray}

Before we go further, let us briefly consider the gauge choice of our monopole $(\vec{A},\Phi)$. Our Lagrangian density is a gauge invariant quantity allowing us to choose some gauge in which to write our monopole fields $(\vec{A},\Phi)$. There are certain advantages and disadvantages to each gauge choice. For instance the choice of gauge where the Higgs field points in just one direction in group space (such as $\hat{e}_3$) is known as the string gauge, and is singular along a Dirac string and at the origin. Another gauge choice, the hedgehog gauge, has the Higgs field pointing in the $\hat{x}$ direction in group space. This field is rotationally invariant under combined rotation and global internal $SU(2)$ transformation. We can transform the monopole from hedgehog gauge to string gauge via a singular gauge transformation. This gauge transformation would change the homotopy class (or winding number) of the monopole. 

The temporal gauge is an incomplete gauge where $A_0=0$. We can consider a time-independent monopole with nonzero $A_0$ as being physically equivalent to a time-dependent monopole in the temporal gauge where we have applied a gauge transformation to compensate for the time dependence of the fields \cite{Shnir2005}. We note that if $A_0\neq 0$ then the monopole has an electric charge in addition to a magnetic charge making it a non-abelian dyon.  $A_0$ would be pointing in the same group space direction as the Higgs field \cite{Shnir2005} allowing us to draw a correspondence between the Higgs field and $A_0$ known as the Julia-Zee correspondence. These gauge choices are surveyed in the pedagogical works of \cite{TongNotes} and \cite{WeinbergYi}.

We choose the gauge of our monopole to be in hedgehog gauge as it will be more convenient to work with.Also since we are interested in static monopoles with just magnetic charge, we assume that all fields are time($x_0$)-independent and that we are in temporal gauge$A_0=0$. This is the static ansatz. We have that
\begin{equation}
E_i=F_{0i}=D_0 A^i - D_i A^0 = \partial_0 A^i = 0,
\end{equation}
and
\begin{equation}
D_0 \Phi = \partial_0 \Phi = 0,
\end{equation}
hence the kinetic energy $T$ vanishes.

We will later relax this assumption when we consider the moduli space of monopoles.
So we are left with
\begin{eqnarray}
E =U &=&\frac{1}{g^2}\left( \left<B_i,B_i\right>+\left<D_i\Phi,D_i\Phi\right>\right),\nonumber\\
	&=& \frac{1}{g^2}\left(\left<D_i \Phi + B_i,D_i \Phi + B_i\right>- 2\left<D_i,B_i\right>\right).
\end{eqnarray}
Now we minimize the energy functional by first setting 
\begin{equation}
D_i \Phi = -B_i.
\label{eqn:Bogomolny}
\end{equation}
These are the Bogomolny equations  \cite{Bogomolny1976}. 

Next we make use of the Bianchi identity and $D_i B_i = -D_i (\epsilon_{ijk} F_{jk}) = 0$ to integrate by parts and compute the $\left<D_i \Phi,B_i\right>$ term.
\begin{eqnarray}
\left<D_i\Phi,B_i\right> &=& \int d^3x Tr(D_i\Phi B_i),\nonumber\\
			&=& \int d^2S^i_\infty Tr(\Phi B_i) - \Sigma^k_n \int d^2S^i_{x\to p_n}Tr(\Phi B_i),
\end{eqnarray}
where $k$ is the number of singularities in $\Phi$ and $p_n$ are the singular points of $\Phi$ and $B^i$ in $\mathbb{R}^3$. The first surface integral is on the surface of a sphere at spatial infinity and the remaining integrals are on the surface of infinitesimally small spheres around the singular points $p_n$. Kronheimer \cite{Kronheimer1985} established the following conditions on $\Phi$ so that these integrals do not diverge
\begin{equation}
\lim_{\vec{x}\to\vec{p}_n} |\vec{x}-\vec{p}_n||\Phi|=\frac{1}{2}l_n,
\end{equation}
and $d(|\vec{x}-\vec{p}_n||\Phi|)$ is bounded as $\vec{x}\to\vec{p}_n$.
We will explore the construction of the dynamics of a monopole with one singularity at the origin in this paper.

Atiyah and Hitchin \cite{AtiyahHitchin1988} provide a useful interpretation of our Yang-Mills-Higgs theory on $\mathbb{R}^3$ with $V(\Phi)=0$ as a Yang-Mills theory on $\mathbb{R}^4$ with a Euclidean metric, identifying $A_0$ with $\Phi$. So in this theory we have the notation
\begin{equation}
D_0 = \partial_0 - i[\Phi,].
\label{eqn:NewD}
\end{equation}
In this interpretation the Bogomolny equations are just the time-independent version of the four-dimensional self-duality equation
\begin{equation}
F_{\mu\nu}=\frac{1}{2} \epsilon_{\mu\nu\rho\sigma} F_{\rho\sigma},
\label{eqn:SD}
\end{equation}
where $F$ is the field strength tensor in our four-dimensional interpretation.

For the gauge group $\mathcal{G}=SU(2)$ there is one simple finite energy solution to the Bogomolny equations called the 't Hooft-Polyakov monopole \cite{tHooft1974} \cite{Polyakov1974}
\begin{eqnarray}
\Phi=\frac{\vec{\sigma}\cdot\vec{r}}{2r^2}\left(2\lambda r\coth{2\lambda r}-1\right),\\
A^i=\left(\frac{2\lambda r}{\sinh{2\lambda r}}-1\right)\frac{(\vec{\sigma}\times\vec{r})_i}{2r^2},
\end{eqnarray}
where $\sigma_i$ are the Pauli matrices that span a representation of $su(2)$. We present the 't Hooft-Polyakov monopole solution because the singular monopole is a non-linear superposition of this basic non-abelian monopole and a Dirac singularity.

For the rest of the paper we set the dimensionless coupling constant $g=1$.

\section{Moduli Space}
As first discussed by Manton \cite{Manton1982} and later by Atiyah and Hitchin \cite{AtiyahHitchin1988}, we can model the low energy dynamics of a monopole by constructing its moduli space. We begin by considering the infinite dimensional configuration space $\mathcal{A}$ of the pair $(A,\Phi)$ modulo ``small" gauge transformations which go to identity on the spatial asymptotic $\mathcal{G}$:  $\mathcal{C}=\mathcal{A}/\mathcal{G}$. On this manifold we have the energy functional $E$ as we defined earlier in equation (\ref{eqn:EnergyFN}). 
The moduli space is the submanifold in $\mathcal{C}$ where $U$ is minimized, and as we have shown $U$ is minimized when the pair $(A,\Phi)$ satisfies the Bogomolny equations. On this submanifold, we make an infinitesimal tangential perturbation to $(A,\Phi)$. We expect $(A',\Phi')$, the newly perturbed configuration, to remain on this submanifold. The motion of the monopole can then be described by the natural metric derived from the kinetic energy term
\begin{equation}
T = \left<E_i,E_i\right>+\left<D_0\Phi,D_0\Phi\right>,
\end{equation}
where
\begin{equation}
E_i=F_{0i} = \partial_0 A^i,
\end{equation}
and
\begin{equation}
D_0\Phi = \partial_0 \Phi,
\end{equation}
using the notation from equation (\ref{eqn:NewD}) and where $A_0=0$. 

Now how do we define the perturbation of $A^i$ and $\Phi$? We look at it from the perspective of a perturbation in the collective coordinates $T^m$. These coordinates specify the different monopole configurations with the same topological charge and will provide a natural set of coordinates for our moduli space. We therefore have that
\begin{eqnarray}
\partial_0 A^i = \delta_m A^i \dot{T}_m,\\
\partial_0 \Phi = \delta_m \Phi \dot{T}_m, 
\end{eqnarray}
where $\delta_m$ is the derivative over all the collective coordinates, $m$ ranges over all the collective coordinates, and where the dot indicates differentiation with respect to $x_0$.  In our case, the $SU(2)$ singular monopole, one of the collective coordinates $m=0$ will correspond to the phase of the monopole and the other three collective coordinates $m=1,2,3$ will correspond to the relative separation of the non-abelian monopole with the singularity.

However we also require that our tangent vectors be transverse to the orbits of the small gauge transformations.  We will call our tangent vector $Z^\mu$.

A small gauge transformation of our pair $(A,\Phi)$ produces a tangent vector $Z'^\mu=D_\mu \Lambda'$ where $\Lambda'$ vanishes at infinity.
So we look for the conditions where $\left<Z^\mu, Z'^\mu\right> = \left<Z^\mu, D_\mu \Lambda'\right>=0$ ensuring transversality to the gauge orbits:
\begin{eqnarray}
0 &=& \int d^3x Tr(Z^\mu D_\mu \Lambda'),\nonumber\\
	&=& \int d^2S^i_\infty Tr(Z^i\Lambda') - \Sigma^k_{n=1}\int d^2S^i_{x\to p_n}Tr(Z^i\Lambda') -\int d^3x Tr(D_\mu Z^\mu \Lambda').\nonumber
\end{eqnarray}
We used integration by parts to go from the first to the second line of the above equation. Recalling that the leading order of $\Lambda'$ vanishes at infinity, the first integral vanishes if $Z^i$ is regular at $\infty$. The $k$ integrals around singular points $p_n$ require the tangent vector $Z^\mu$ to be sufficiently smooth at the singularities. The last integral imposes the condition
\begin{equation}
D_\mu Z^\mu =0,
\label{eqn:BGC}
\end{equation}
making the tangent vector $Z^\mu$ a zero mode of adjoint covariant derivative $D_\mu$.
We therefore define our tangent vectors $Z^\mu=\delta A^\mu+D_\mu \Lambda$ where we have added a compensating gauge term $D_\mu \Lambda$ such that $Z^\mu$ satisfies eq. (\ref{eqn:BGC}).
By differentiating the Bogomolny equations we also have that the tangent vector $Z^m_\mu$ satisfies the linearized Bogomolny equations.
\begin{equation}
D_i Z^j - D_j Z^i = D_0 Z^k - D_k Z^0, 
\label{eqn:LinBog}
\end{equation}
where $i,j,k$ are an even permutation of $1,2,3$.

We can then write the appropriate form of the kinetic energy as
\begin{equation}
T = \left<Z^i,Z^i\right>+\left<Z^0,Z^0\right> = \left<Z^\mu,Z^\mu\right>,
\end{equation}
where $\mu=0,1,2,3$. 

We can write the tangent vector as $Z^\mu = Z_m^\mu \dot{T}_m$ so that the kinetic energy is now
\begin{equation}
T= \left<Z_m^\mu,Z_n^\mu\right> \dot{T}_m\dot{T}_n,
\end{equation}
We define the metric on our moduli space as
\begin{equation}
g_{mn} = \left<Z_m^\mu,Z_n^\mu\right>.
\label{eqn:metricformula}
\end{equation}
A useful physical interpretation of our derivation of the metric on the moduli space is as follows. We start with a static monopole localized in $\mathbb{R}^3$ and its position and phase specified by the collective coordinates $T^m$. We then make these collective coordinates time-dependent $T^m=T^m(x_0)$. For slowly changing collective coordinates, the monopole changes with time but the field configuration of the monopole remains close to our manifold of static monopole solutions. This means that we are moving from one point to another point on the moduli space parameterized by the collective coordinates $T^m$. Therefore we can naturally derive a metric from the kinetic energy term generated by the time dependence of $T^m$ as the metric on the moduli space. This is the low energy moduli space approximation as first described by  \cite{Manton1982}.

Also we can now think of the metric as computing the overlap of integrable zero modes of $D_\mu$ in the background of the monopole $(A,\Phi)$.

\section{Construction of the $SU(2)$ Singular Monopole}

We now go on to construct monopoles by using an extension of the ADHM construction \cite{ADHM1978} of instantons developed by Nahm \cite{Nahm1980a} \cite{Nahm1980b} \cite{Nahm1981}. The Nahm transform was formalized by Hitchin \cite{Hitchin1983} and generalized to classical groups by Hurtubise and Murray \cite{HurtubiseMurray1989}.

We directly follow B. Durcan's thesis \cite{Durcan2007} to compute the explicit form of the $SU(2)$ singular monopole using the Nahm transform.

The Nahm equations are the self-duality equations eq.(\ref{eqn:SD}) when the connection is $x_1,x_2,x_3$ invariant. We can directly construct solutions to the Bogomolny equations by using the solutions of the Nahm equations.

First we find solutions to the Nahm equations
\begin{equation}
\partial_s T_i(s)-i[T_0(s),T_i(s)]=-i[T_j(s),T_k(s)],
\end{equation}
where $i,j,k$ are an even permutation of $1,2,3$ and $T^\mu$ are Hermitian $k\times k$ ($k\in\mathbb{N}$) matrices. The Nahm data of rank $k$ are used to construct solutions to the Bogomolny equations in non-abelian gauge groups with non-abelian topological charge $k$. The Nahm data can be used as the collective coordinates of the resulting monopole. We note explicitly the gauge transformation of $T^\mu$
\begin{eqnarray}
T^0\to g^{-1}T^0g+ig^{-1}\partial_s g,\\
T^i\to g^{-1}T^ig ,
\label{eqn:NahmGauge}
\end{eqnarray}
where $g(s)\in U(k)$.

For $k=1$ this is just the $U(1)$ abelian gauge transformation. For the rest of this paper we will restrict our attention to Nahm data of rank $k=1$ and consequently to monopoles of non-abelian charge $1$. With $k=1$ we have that the commutators vanish in the Nahm equations leaving $\partial_s T_i(s) = 0$ which means that $T_i(s)$ is constant.

Our Nahm data for the $SU(2)$ singular monopole are defined as a piecewise function over the real line $s\in \mathbb{R}$ as follows
\begin{eqnarray}
\vec{T}=\vec{T}_M(s)&=&\vec{T}_{'tHP}\in\mathbb{R}^3 \text{for} s\in(-\lambda,\lambda),\\
\vec{T}=\vec{T}_R(s)=\vec{T}_L(s)&=&\vec{T}_D\in\mathbb{R}^3 \text{for} |s|>\lambda,
\end{eqnarray}
where we labeled these intervals as follows: $L$ is $(-\infty,-\lambda)$, $R$ is $(\lambda,\infty)$, and $M$ is $(-\lambda,\lambda)$.

For our convenience we define the following quantities
\begin{eqnarray}
\vec{r}=\vec{x}-\vec{T}_{'tHP},\\
\vec{z}=\vec{x}-\vec{T}_D,\\
\vec{d}=\vec{T}_{'tHP}-\vec{T}_D=\vec{z}-\vec{r},
\end{eqnarray}
$\vec{T}_D$ is the position of the Dirac singularity in our final monopole solution and $\vec{T}_{'tHP}$ is the position of the non-abelian 't Hooft-Polyakov type monopole.
We translate our system so that $\vec{T}_D$ is at the origin: $\vec{T}_D=0$ and $\vec{T}_{'tHP}=\vec{d}$. We rewrite the Nahm data as $\vec{T}(s)=\Theta'(s)\vec{d}$ where $\Theta '(s)=\Theta(s+\lambda)-\Theta(s-\lambda)$  with $\Theta$ as the Heaviside step function. We can gauge $T_0$ away using eq. (\ref{eqn:NahmGauge}).

We now discuss the boundary conditions and boundary data of the Nahm data. We are particularly interested in the boundary conditions of the Nahm data for intervals where the rank of the Nahm data is $k=1$ and does not change over the boundary. With our singular monopole there is a discontinuity in $T^i$ at boundary points $s=\pm \lambda$. If we have two adjacent intervals with the same rank Nahm data we must introduce jumping data $f_\pm$ at the boundaries between intervals. These jumping data for $k=1$ take the form of a set of $2\times 1$ complex spinors. For a further pedagogical introduction to jumping data see \cite{HurtubiseMurray1989} and \cite{ChenWeinberg2003}.

Due to these discontinuities we modify the Nahm equations to be as follows
\begin{equation}
\frac{dT_i}{ds}-i[T_0,T_i]=-i[T_j,T_k]-\frac{1}{2}\delta(s-\lambda)f_+^\dagger\sigma_if_+-\frac{1}{2}\delta(s+\lambda)f_-^\dagger\sigma_if_-.
\end{equation}

Let us briefly review some of the properties of the spinors $f_+$ and $f_-$. For $k=1$ the commutators vanish and we take the integral of the Nahm equations over a vicinity of $s=\pm\lambda$ and find $\pm d^i = \frac{1}{2}f^\dagger_\pm\sigma_if_\pm$. This result implies that $f^\dagger_\pm \sigma_i f_\pm=\pm 2d_i$, $f_\pm^\dagger f_\pm=2d$, and $f_\pm f^\dagger_\pm = d I_{2\times 2} \pm \vec{\sigma}\cdot\vec{d}$ with $d=|\vec{d}|$. We also have that $\vec{\sigma}\cdot\vec{d} f_\pm=\pm d f_\pm$. These spinors can be chosen up to an overall $U(1)$ phase. A gauge transformation on the Nahm data will transform the spinor $f_\pm$ as $f_\pm\to g(\pm\lambda)f_\pm$. 

We define a $2\times 2$ matrix $\varphi=\begin{pmatrix} f^\dagger_+ \\ f^\dagger_- \end{pmatrix}$ and find that $Tr(\varphi^\dagger \varphi)=4d$.  We use $\varphi$ to simplify the notation of our monopole construction.

We define the following quantities
\begin{eqnarray}
\hat{\varphi} \equiv\delta(s-\lambda) P_1 \varphi+\delta(s+\lambda)P_2\varphi,\\
\slashed{D}^\dagger_x = -(\partial_s+ix^0)I_{2\times 2}-\sigma_j(T^j-x^j),\\
\Delta=\begin{pmatrix}\Delta_+\\ \Delta_- \end{pmatrix},\\
\hat{\Delta}=\delta(s-\lambda)P_1\Delta+\delta(s+\lambda)P_2\Delta,\\
\end{eqnarray}
and the $2\times 2$ projection matrices  $P_1= \frac{1+\sigma_3}{2}$ and $P_2=\frac{1-\sigma_3}{2}$. We note that $\Delta_{\pm}$ are $1\times 2$ row vectors.
We write the twisted adjoint Weyl operator as 
\begin{equation}
\hat{D}^\dagger= \begin{pmatrix} \hat{\varphi}^\dagger & \slashed{D}^\dagger_x \end{pmatrix}.
\end{equation}

We then write the Weyl equation for the singular monopole 
\begin{eqnarray}
\hat{D}^\dagger v &=& \begin{pmatrix} \hat{\varphi}^\dagger & \slashed{D}^\dagger_x \end{pmatrix} \begin{pmatrix} \hat{\Delta} \\ \psi \end{pmatrix},\nonumber\\
&=& (-(\partial_s+ix^0)I_{2\times 2}-\sigma_j\otimes(T^j(s)-x^j))\psi\\ \nonumber
&&+\delta(s-\lambda)f_+\Delta_++\delta(s+\lambda)f_-\Delta_-=0,
\label{eqn:WeylEq}
\end{eqnarray}
where $v=\begin{pmatrix} \hat{\Delta} \\ \psi \end{pmatrix}$ is in the kernel of the fundamental Weyl operator.

In our case for $SU(2)$ there are two independent solutions to $\slashed D^\dagger_x \psi=0$ over $\mathcal{I}$ so $\psi$ is a $2\times 2$ matrix with columns forming an orthonormal basis of all solutions.
 
We adopt the instanton notation, used in \cite{Dorey1996} \cite{Kraan}, for the rest of the paper to describe our monopole constructions. In this notation $v^\dagger v = \Delta^\dagger_+\Delta_++\Delta^\dagger_-\Delta_-+\int^\infty_{-\infty} ds \psi^\dagger(s)\psi(s)$. 
We choose normalized $v$ such that $v^\dagger v=1_{2\times 2}$.

The following solutions to the Weyl equation are found in \cite{Durcan2007}:

\begin{eqnarray}
\psi_L(s)&=& e^{(\vec{\sigma}\cdot\vec{z}-ix_0 )(s+\lambda)}\frac{\zeta_+f^\dagger_+}{f^\dagger_+\zeta_+}\psi_{'tHP}(-\lambda)N,\\
\psi_M(s) &=& e^{-ix_0 s}\psi_{'tHP}(s)N,\\
\psi_R(s) &=& e^{(\vec{\sigma}\cdot\vec{z}-ix_0)(s-\lambda)}\frac{\zeta_-f^\dagger_-}{f^\dagger_-\zeta_-}\psi_{'tHP}(\lambda)N,\\
\Delta_{-} &=& e^{ix_0\lambda}\frac{\zeta^\dagger_-}{\zeta^\dagger_-f_-}\psi_{'tHP}(-\lambda)N,\\
\Delta_{+} &=& -e^{-ix_0\lambda}\frac{\zeta^\dagger_+}{\zeta^\dagger_+f_+}\psi_{'tHP}(\lambda)N,
\label{eqn:PsiDef}
\end{eqnarray}
where $\zeta_\pm$ are spinors with the same properties we described for $f_\pm$ where $\vec{\sigma}\cdot\vec{z}\zeta_\pm=\pm z\zeta_\pm$, $\zeta^\dagger_\pm\zeta_\pm=2z$, 
and
\begin{eqnarray}
\psi_{'tHP}(s) &=& \sqrt{\frac{r}{\sinh (2\lambda r)}}e^{\vec{\sigma}\cdot\vec{r}s},\\
a &=& z+d,\\
\mathcal{D} &=& a^2-r^2 = 2zd+2\vec{z}\cdot\vec{d},\\
\mathcal{K} &=& (a^2+r^2)\cosh (2\lambda r)+2ra\sinh (2\lambda r),\\
\mathcal{L} &=& (a^2+r^2)\sinh (2\lambda r)+2ra\cosh (2\lambda r),\\
N &=& \sqrt{\frac{\mathcal{D}}{a^2+r^2+2ra\coth (2\lambda r)}}=\sqrt{\frac{\mathcal{D}\sinh (2\lambda r)}{\mathcal{L}}}. 
\end{eqnarray}
We note that $\psi_{'tHP}$ is the Weyl solution we would use to construct the 't Hooft-Polyakov monopole.

We define the Higgs field and connection as 
\begin{equation}
\Phi=i v^\dagger \partial_0 v= \lambda \Delta^\dagger_{+}\Delta_{+} -\lambda \Delta^\dagger_{-}\Delta_{-}+\int_{-\infty}^\infty ds \psi^\dagger(s) s \psi(s),
\end{equation}
 and 
 \begin{equation}
 A^i = iv^\dagger \partial_i v =i\Delta^\dagger_{+\lambda}\partial_i \Delta_{+} +i\Delta^\dagger_{-}\partial_i\Delta_{-}+i\int_{-\infty}^\infty ds \psi^\dagger(s) \partial_i \psi(s).
 \end{equation}

The Higgs field and connection are as follows
\begin{eqnarray}
\Phi 	&=&	\vec{\sigma}\cdot\hat{r}\left(\left(\lambda+\frac{1}{2z}\right)\frac{\mathcal{K}}{\mathcal{L}}-\frac{1}{2r}\right)-\frac{r}{z\mathcal{L}}\vec{\sigma}\cdot\vec{d}_\perp,\\
A_i	&=&	\left(\left(\lambda+\frac{a}{\mathcal{D}}\right)\frac{\mathcal{D}}{\mathcal{L}}-\frac{1}{2r}\right)\frac{(\vec{\sigma}\times\vec{r})_i}{r}\nonumber\\
&&-\frac{r}{\mathcal{L}}\left(\frac{(\vec{\sigma}\times\vec{z})_i}{z}+\left(\frac{\mathcal{K}}{\mathcal{D}}-1\right)\frac{(\vec{r}\times\vec{z})_i}{rz}\vec{\sigma}\cdot\hat{r}\right).
\label{eqn:APhi}
\end{eqnarray}

One can show that this solution satisfies the Bogomolny equations and have a non-abelian charge $1$. The 1-dimensional Green's function of this singular monopole as derived by Durcan in \cite{Durcan2007} can be found in Appendix A. The parameter $\lambda$ is of dimension $1/l$. The ``radius'' of the non-abelian monopole core is inversely proportional to $\lambda$, $R\sim1/\lambda$, and the mass is proportional to $\lambda$, $M\sim\lambda/g^2$, giving mass the appropriate dimensions.

It was shown in \cite{Durcan2007} and can be seen from the above solution that $\vec{T}_D$ corresponds to the location of a Dirac monopole type singularity and $\vec{T}_{'tHP}$ corresponds to the location of the non-abelian 't Hooft-Polyakov type monopole. For intervals which are semi-infinite, such as $L$ and $R$, we find a pole/singularity with abelian charge in our Higgs field. The gauge fields near $\vec{T}_D$ approach that of a Dirac monopole in some $U(1)$ subgroup generator of $SU(2)$.



\chapter{Translational and Phase Zero Modes} 
\label{Chapter2}
\lhead{Chapter 2. \emph{Translational and Phase Zero Modes}} 

\section{Zero Mode Construction}
Our first ``attack'" in computing the tangent vectors we would need to calculate the moduli space metric of the singular monopole is to compute $\delta_m A^i$ and $\delta_m \Phi$ using our previously constructed singular monopole pair $(A,\Phi)$ in equation (\ref{eqn:APhi}), and where $\delta_m \equiv  \frac{\partial}{\partial T^m} \equiv \frac{\partial}{\partial d^m}$ and $m=0,1,2,3$ over the collective coordinates $T^m=d^m$. 

\begin{eqnarray}
\delta_0 \Phi &=& 0,\\
\delta_q \Phi &=& \frac{1}{\mathcal{L}}(\vec{\sigma}\cdot\hat{r}\left\{ \frac{r_q}{z}+2(\hat{z}_q+\hat{d}_q)\left(\lambda+\frac{1}{2z}\right)r\frac{\mathcal{D}}{\mathcal{L}}-\frac{\hat{z}_q}{z^2}\mathcal{K}\right\}\nonumber\\
&& +\frac{r}{z}\vec{\sigma}\cdot\vec{d}_\perp\left\{\frac{\hat{z}_q}{z^2}-\frac{2}{\mathcal{L}}(\hat{z}_q+\hat{d}_q)N\right\}-\frac{r}{z}\sigma_q),
\end{eqnarray}

\begin{eqnarray}
\delta_0A_i &=& 0,\\
\delta_q A_i &=& \frac{1}{\mathcal{L}}((\vec{\sigma}\times\vec{r})_i\left \{(\hat{z}_q+\hat{d}_q)\left[\left(2\lambda+\frac{2a}{\mathcal{D}}\right)\left(a-\frac{N}{\mathcal{L}}\right)-\frac{2r}{z\mathcal{L}}N-\frac{a^2+r^2}{\mathcal{D}}\right]+r\frac{\hat{z}_q}{z^2}\right \}\nonumber\\
&&+(\vec{\sigma}\times\vec{d})_i\left \{r\frac{\hat{z}_q}{z^2}-\frac{2r}{z\mathcal{L}}N(\hat{z}_q+\hat{d}_q)\right \}\nonumber\\
&&-\vec{\sigma}\cdot\hat{r}\{ \left( \frac{\mathcal{K}}{\mathcal{D}}-1\right) \left( (\hat{r}\times\hat{z})_i\left(\frac{2r(\hat{z}_q+\hat{d}_q)}{\mathcal{L}}N-\frac{\hat{z}_q}{z^2}\right)+\frac{r}{z}(\hat{r}\times\hat{e}_q)_i\right)\nonumber\\
&&+\frac{2r(\hat{z}_q+\hat{d}_q)}{\mathcal{D}}(\hat{r}\times\hat{z})_i\left(M-a\frac{\mathcal{K}}{\mathcal{D}}\right)\}-\frac{r}{z}(\vec{\sigma}\times\hat{e}_q)_i),
\label{eqn:TanVectors}
\end{eqnarray}
where $C=\cosh 2\lambda r$, $S=\sinh 2\lambda r$, $M=aC+rS$, $N=aS+rC$, $\hat{e}_q$ is the unit vector in the $q$ direction, and $q=1,2,3$ over the translational collective coordinates.

We impose the condition that our tangent vectors are orthogonal to small gauge transformations, as in eq. (\ref{eqn:BGC}). This implies that if $\delta_m A^\mu$ is the tangent vector, we must check to see if it satisfies
\begin{equation}
D_\mu \delta_m A^\mu=0.
\label{eqn:BGC2}
\end{equation}
However our current translational tangent vectors do not satisfy this condition without some compensating gauge term. We will also need to find a non-trivial gauge transformation $\Lambda_m$ so that $D_\mu \left( \delta_m A^\mu +D_\mu \Lambda_m\right)=0$.

We can add a gauge term to $\delta_m A^\mu$ while still satisfying the linearized Bogomolny equations eq. (\ref{eqn:LinBog}).
So we naively write our tangent vectors as
\begin{equation}
Z^\mu_m = \delta_m A^\mu +D_\mu \Lambda_m.
\end{equation}

Our goal is to now see if we can find $\Lambda_m$ such that  some $Z^\mu_m$ satisfies the background gauge condition eq. (\ref{eqn:BGC}): $D_\mu(D_\mu \Lambda_m)=-D_\mu (\delta_m A^\mu)$.

We start by making an ansatz, as in  \cite{LeeYi1998} \cite{Dorey1996}, that 
\begin{equation}
\Lambda_m = v^\dagger Q_m v,
\label{eqn:LambdaDef}
\end{equation}
 where $Q_m=\begin{pmatrix} \chi_m \sigma_3 & 0 \\ 0 & p_m(s)I_{2\times 2}\end{pmatrix}$ where $p^m(s)\in\mathbb{R}$ and $\partial_\mu Q_m=0$. We note that relaxing the condition $\partial_\mu Q_m=0$ can yield to a slightly alternative derivation of the compensating gauge term than described here.

Let us first calculate $D_\mu \Lambda_m$ 
\begin{eqnarray}
D_\mu \Lambda_m &=& \partial_\mu (v^\dagger Q_m v)-i[ iv^\dagger\partial_\mu v,v^\dagger Q_m v],\\
&=& \partial_\mu v^\dagger Q_m v + v^\dagger Q_m \partial_\mu v-\partial_\mu v^\dagger v v^\dagger Q_m v-v^\dagger Q_m v v^\dagger\partial_\mu,\\
&=& \partial_\mu v^\dagger \hat{D} F \hat{D}^\dagger Q_m v + v^\dagger Q_m \hat{D} F \hat{D}^\dagger \partial_\mu v.
\end{eqnarray}
Here we have used that $v$ are our previously found solutions eq.(\ref{eqn:PsiDef}) to the Weyl equation eq.(\ref{eqn:WeylEq}), $v^\dagger v=I_{2\times 2}$ and the projection operator $P=1-v v^\dagger = \hat{D} F \hat{D}^\dagger$ where $F(s,t)$ is the Green's function of the covariant Laplacian: $\hat{D}^\dagger \hat{D} F(s,t) = \delta(s-t)$. 

We must now compute $\hat{D}^\dagger Q_m v$.
\begin{eqnarray}
\hat{D}^\dagger Q_m v &=& \begin{pmatrix} \hat{\varphi}^\dagger & \slashed{D}_x^\dagger \end{pmatrix} \begin{pmatrix} \chi_m \sigma_3 & 0 \\ 0 & p_m(s)I_{2\times 2}\end{pmatrix}\begin{pmatrix} \hat{\Delta} \\ \psi \end{pmatrix},\nonumber\\
&=& \begin{pmatrix} \hat{\varphi}^\dagger & \slashed{D}_x^\dagger \end{pmatrix} \begin{pmatrix} \chi_m\sigma_3\hat{\Delta} \\ p_m(s)\psi \end{pmatrix},\nonumber\\
&=& \hat{\varphi}^\dagger \chi_m \sigma_3\hat{\Delta} + \slashed{D}_x^\dagger (p_m(s)\psi),\nonumber\\
&=& \hat{\varphi}^\dagger \chi_m \sigma_3 \hat{\Delta} -\partial_sp_m(s) \psi +p_m(s)\slashed{D}^\dagger_x \psi,\nonumber\\
&=& \hat{\varphi}^\dagger \chi_m \sigma_3 \hat{\Delta} -\partial_sp_m(s)\psi-p_m(s)\hat{\varphi}^\dagger\Delta,\nonumber\\
&=& \begin{pmatrix} \hat{\varphi}^\dagger(\chi_m\sigma_3-p_m(s)) & -\partial_sp_m(s)\end{pmatrix} v,
\end{eqnarray}
where we use $\hat{D}^\dagger v=0 \to \hat{\varphi}^\dagger \hat{\Delta}=-\slashed D^\dagger_x \psi$ to go from the 4th to the 5th line of the derivation.
We define 
\begin{equation}
G_m = \begin{pmatrix} (\chi_m\sigma_3-p_m(s))\hat{\varphi} \\ -\partial_s p_m(s).\end{pmatrix}.
\label{eqn:Gm}
\end{equation}

To further simplify $D_\mu \Lambda_m$ we use the following simplification

\begin{eqnarray}
\hat{D}^\dagger \partial_\mu v &=& -(\partial_\mu \hat{D}^\dagger) v, \nonumber\\
&=& -\begin{pmatrix} \partial_\mu \hat{\varphi}^\dagger & \partial_\mu \slashed D^\dagger_x\end{pmatrix} \begin{pmatrix} \hat{\Delta} \\ \psi \end{pmatrix}, \nonumber\\
&=& -\begin{pmatrix} 0 & \bar{\sigma}_\mu \end{pmatrix} \begin{pmatrix} \hat{\Delta} \\ \psi \end{pmatrix},\nonumber\\
&=& -\bar{\sigma}_\mu \psi,
\end{eqnarray}

where we defined our quaternionic basis as $\sigma_\mu=(i I_{2\times 2},\sigma_i)$ and $\bar{\sigma}_\mu=(-i I_{2\times 2},\sigma_i)$.

We can then write $D_\mu \Lambda^m$ as
\begin{eqnarray}
D_\mu \Lambda^m &=& -\psi^\dagger \sigma_\mu F G^\dagger_m v-v^\dagger G_m F \bar{\sigma}_\mu \psi, \nonumber\\
&=& -\int ds dt \psi^\dagger(s) \sigma_\mu F(s,t) \left( (\chi_m\sigma_3-p_m(s))\hat{\varphi}^\dagger\hat{\Delta}-\partial_s p_m(s)\psi(s)\right) - h.c.,\nonumber
\end{eqnarray}
where $h.c.$ stands for the Hermitian conjugate of the term preceding it.

Now we would like to see if we can get $\delta_m A^\mu$ into a similar form, following a similar derivation for instantons in  \cite{CGTO1979} and  \cite{Dorey1996}.

We start with the definition of $A^\mu=iv^\dagger\partial_\mu v$. We then differentiate $A^\mu$ by $\delta_m$ and find 
\begin{eqnarray}
\delta_m A^\mu &=& i\delta_m v^\dagger \partial_\mu v+iv^\dagger\partial_\mu \delta_m v,\nonumber\\
&=&i\delta_m v^\dagger \partial_\mu v+i\partial_\mu (v^\dagger\delta_m v)-i\partial_\mu v^\dagger \delta_m v,\nonumber\\
&=& D_\mu(iv^\dagger\delta_m v)-\psi^\dagger \sigma_\mu F (i\delta_m \hat{D}^\dagger)v -v^\dagger(-i\delta_m\hat{D})F\bar{\sigma}_\mu\psi.
\end{eqnarray}
We used the projection operator $P=1-vv^\dagger=\hat{D}F\hat{D}^\dagger$ to go from the 2nd to the 3rd line of the derivation and we used $\hat{D}^\dagger\partial_\mu v=-\bar{\sigma}_\mu \psi$.

We define $C_m=-i\delta_m \hat{D}= \begin{pmatrix} -i \delta_m \hat{\varphi} \\ i\sigma_m \end{pmatrix}$. We combine $C_m$ and $G_m$, $H_m=C_m+G_m$.
We define our zero mode $Z^\mu_m$ as 
\begin{equation}
Z^\mu_m =\delta_m A^\mu + D_\mu( \Lambda_m-iv^\dagger \delta_m v) = \delta_m A^\mu +D_\mu(\Omega_m),
\end{equation}
where we define the gauge term $\Omega^m=\Lambda_m-iv^\dagger \delta_m v$.
We can then redefine $Z^\mu_m$ as
\begin{equation}
Z^\mu_m = v^\dagger H_m F \bar{\sigma}_\mu \psi + \psi^\dagger \sigma_\mu F H^\dagger_m v.
\label{eqn:ZDef}
\end{equation}

We would like to see if there was some way we can determine $\chi^m$ and $p^m(s)$ such that $Z^\mu_m$ satisifies eq. (\ref{eqn:BGC2}).
We first take the adjoint covariant derivative of $Z^\mu_m$:
\begin{equation}
D_\mu Z^\mu_m = \psi^\dagger F \sigma_\mu (H^\dagger_m\hat{D}+\hat{D}^\dagger H_m).\bar{\sigma}_\mu F \psi = 0
\end{equation}
To satisfy this equation we require that 
\begin{equation}
Tr(\hat{D}^\dagger H_m+H^\dagger_m\hat{D})=0.
\label{eqn:BGCGeneral}
\end{equation}

We adopt the notation used in \cite{Kraan} and \cite{KraanVanBaal1998} to simplify some of our further calculations, 
\begin{equation}
H =H_m (dT)^m = \begin{pmatrix} \hat{c} \\ \hat{Y} \end{pmatrix},
\label{eqn:HDef}
\end{equation}
where 
\begin{eqnarray}
\hat{c} &=& \delta(s-\lambda)P_1(-i\delta\varphi+\chi\varphi-p(\lambda)\varphi)+\delta(s+\lambda)P_2(-i\delta \varphi-\chi\varphi-p(-\lambda)\varphi),\nonumber\\
\hat{Y}&=&(i\vec{\sigma}\cdot d\vec{T}-\partial_s p(s))\Theta'(s),
\end{eqnarray}
where $p(s)=p_m (dT)^m$,  and $\chi=\chi_m(dT)^m$. Once we construct our gauge compensating terms $\Lambda^m$ we can determine what we must set as the value of $\chi_m$.

We now insert these definitions eq. (\ref{eqn:HDef})  into eq. (\ref{eqn:BGCGeneral}) to derive an equation for $p(s)$:
\begin{eqnarray}
Tr(\hat{D}^\dagger H+H^\dagger \hat{D}) &=& \partial^2_s p(s)\nonumber\\
&& +\delta(s+\lambda)((-\chi-p(-\lambda))Tr(\varphi^\dagger P_2\varphi)-\frac{i}{2}Tr(\varphi^\dagger P_2\delta\varphi-\delta\varphi^\dagger P_2\varphi))\nonumber\\
&&+\delta(s-\lambda)((\chi-p(\lambda))Tr(\varphi^\dagger P_1\varphi)-\frac{i}{2}Tr(\varphi^\dagger P_1\delta\varphi-\delta\varphi^\dagger P_1 \varphi)),\nonumber\\
&=& 0.
\label{eqn:NahmBGC}
\end{eqnarray}

We make an important note here that the vector $H$ is the tangent vector over Nahm data. We will use the tangent vector $H$ in Chapter 4 to directly compute the moduli space metric over the Nahm data.


After computation of the trace terms (see Appendix A, eq.(\ref{eqn:trterms})), eq. (\ref{eqn:NahmBGC}) becomes
\begin{eqnarray}
0 &=& \partial^2_s p(s)+\delta(s+\lambda)(-2dp(-\lambda)-2d(dT^0+\vec{\omega}\cdot d\vec{T}+\vec{\chi}\cdot d\vec{T}))\nonumber\\
&&+\delta(s-\lambda)(-2dp(\lambda)+2d(dT^0+\vec{\omega}\cdot d\vec{T}+\vec{\chi}\cdot d\vec{T})),
\label{eqn:NahmBGC2}
\end{eqnarray}
where we used the results from \cite{GibbonsEtAl} to simplify the quantities we traced over in eq. (\ref{eqn:NahmBGC}).
The vector $\vec{\omega}$ is the Dirac monopole connection \cite{Dirac1931} which is the solution to the $U(1)$ Bogomolny equations $\vec{\nabla}_d\times\vec{\omega}=\vec{\nabla}_d V$ with potential $V=\lambda+\frac{1}{2d}$. 

From this equation we can see that away from $s=\pm\lambda$ the second derivative of $p(s)$ is zero which means that it is linear. We also see that when we integrate over a small neighborhood of either $s=+\lambda$ or $s=-\lambda$,
\begin{eqnarray}
-\partial_s p(s)|^{-\lambda+}_{-\lambda-}&=&\partial_s p(-\lambda-)-\partial_s p(-\lambda+)=-2dp(-\lambda)-2d(dT^0+\vec{\omega}\cdot d\vec{T}-\vec{\chi}\cdot d\vec{T})),\nonumber\\
-\partial_s p(s)|^{+\lambda+}_{+\lambda-}&=&\partial_s p(\lambda-)-\partial_s p(\lambda+)=-2dp(\lambda)+2d(dT^0+\vec{\omega}\cdot d\vec{T}-\vec{\chi}\cdot d\vec{T})),\nonumber\\
\label{eqn:pdisc}
\end{eqnarray}
where there are discontinuities at $s=\pm\lambda$ in the first derivative of $p(s)$. This tells us that $p(s)$ is a piecewise linear function.
Given these conditions and eqs. (\ref{eqn:pdisc}) we can find $p(s)$ to be 
\begin{eqnarray}
p_M(s)=\frac{s}{V}(dT^0+\vec{\omega}\cdot d\vec{T}+\vec{\chi}\cdot d\vec{T})),\nonumber\\
p_R(s>\lambda)=\frac{\lambda}{V}(dT^0+\vec{\omega}\cdot d\vec{T}+\vec{\chi}\cdot d\vec{T})),\nonumber\\
p_L(s<-\lambda)=-\frac{\lambda}{V}(dT^0+\vec{\omega}\cdot d\vec{T}+\vec{\chi}\cdot d\vec{T})).
\end{eqnarray}

\subsection{Phase Zero Mode Construction}
We start with the gauge zero mode $Z^\mu_0 = D_\mu \Lambda_0$ which is the zero mode associated with ``large" gauge transformation, which ``rotates" the monopole's framing at infinity in the unbroken $U(1)$ subgroup of $SU(2)$.

We note that $\Lambda_0$ is the adjoint solution to the covariant Laplacian $D_\mu D_\mu \Lambda_0=0$ using $p_0=\frac{s}{V}$ and $\chi_0=1$:
\begin{equation}
Q_0(s)=\begin{pmatrix} \sigma_3 & 0 \\ 0 & \frac{s}{V}\end{pmatrix}.
\label{eqn:Q0}
\end{equation}

We use our previous equation (\ref{eqn:LambdaDef}) for $\Lambda^0$
\begin{eqnarray}
\Lambda^0 &=& \int ds \begin{pmatrix} \hat{\Delta}^\dagger & \psi^\dagger(s) \end{pmatrix} \begin{pmatrix} \sigma_3 & 0_{2\times 2} \\ 0_{2\times 2} & s\frac{1}{V} \end{pmatrix} \begin{pmatrix} \hat{\Delta} \\ \psi(s) \end{pmatrix}, \nonumber\\
&=& \int ds \Delta^\dagger_+\Delta_+-\Delta^\dagger_-\Delta_-\nonumber\\
&&+\int_{-\infty}^{-\lambda} ds \psi_R^\dagger(s) p_0(s>\lambda) \psi_R(s)+\int_\lambda^\infty ds \psi_L^\dagger(s) p_0(s<-\lambda) \psi_L(s)\nonumber\\
&&+\int_{-\lambda}^{\lambda} ds \psi_M^\dagger(s) p_0(s\in(-\lambda,\lambda)) \psi_M(s), \nonumber \\
&=& \frac{r}{\mathcal{L}}(\vec{\sigma}\cdot\hat{r}(2z\sinh (2\lambda r)+2\cosh (2\lambda r)(\vec{z}\cdot\hat{r})-2(\vec{z}\cdot\hat{r})+2r\nonumber \\
&& +\frac{\lambda}{V}(2d\sinh (2\lambda r)-2\cosh (2\lambda r)(\hat{r}\cdot\vec{d})+2(\hat{r}\cdot\vec{d})\nonumber\\
&&+\mathcal{D}\frac{\sinh (2\lambda r)}{2\lambda r^2}\left(2\lambda r\coth (2\lambda r)-1\right))+\frac{1}{Vd}\vec{\sigma}\cdot\vec{d}).
\label{eqn:L0}
\end{eqnarray}

We find the following leading and subleading order for the asymptotic expansion (where $r\to\infty$) of $\Lambda_0$.
\begin{equation}
\Lambda_0 = \vec{\sigma}\cdot\hat{r}\left(1-\frac{1}{2V}\frac{1}{r}+O(r^{-2})\right).
\label{eqn:L0Asymp}
\end{equation}

Next we calculate $\Lambda_q$ using $p_q(s)$. We can split $p_q(s)$ (where $p(s)=p_m(s) dT^m$) into two parts: $p_q(s)=\frac{s}{V}\omega_q+\chi_qp_0(s)$ where $q=1,2,3$ over the translational collective coordinates.  Using this form of $p_q(s)$ we can split the construction of $\Lambda_q$ into two simple computations by splitting $Q_q$ into two parts: 
\begin{equation}
Q_q = \begin{pmatrix} \chi^q \sigma_3 & 0_{2\times 2} \\ 0_{2\times 2} & s\frac{\omega_q}{V} + \chi^q p_0(s) \end{pmatrix} = Q^*_q + \chi_q Q_0,
\end{equation}
where $Q^*_q=\begin{pmatrix} 0_{2\times 2} & 0_{2\times 2} \\ 0_{2\times 2} & s\frac{\omega_q}{V}\end{pmatrix}$.

We plug this into (\ref{eqn:LambdaDef}) and find $\Lambda_q=v^\dagger Q^*_qv+\chi_qv^\dagger Q_0v=\Lambda_q^*+\chi_q\Lambda_0$. From this and the asymptotic expansion eq. (\ref{eqn:L0Asymp}) we see that a non-zero $\chi_q$ changes the framing of the monopole at infinity. Since this direction from $\chi_q$ corresponds to a change in phase, and we seek to only study the translational modes, we are looking for no change in framing from our gauge term $\Lambda^q$. Therefore we must have that $\chi_q=0$. So $\Lambda_q = \Lambda_q^*$.
That just leaves us to compute $\Lambda^q$ using $Q^*_q$:

\begin{eqnarray}
\Lambda_q&=& \int ds \begin{pmatrix} \hat{\Delta}^\dagger & \psi^\dagger(s) \end{pmatrix} \begin{pmatrix} 0_{2\times 2} & 0_{2\times 2}\\ 0_{2\times 2} & s\frac{\omega_q}{V} I_{2\times 2}\end{pmatrix} \begin{pmatrix}\hat{\Delta} \\ \psi(s) \end{pmatrix}, \nonumber\\
&=&  \int_\lambda^\infty ds \lambda\frac{\omega_q}{V}\psi_R^\dagger(s) \psi_R(s)-\int_{-\infty}^{-\lambda} ds \lambda \frac{\omega_q}{V} \psi_L^\dagger(s) \psi_L(s)\nonumber\\
&&+\int_{-\lambda}^{\lambda} ds \frac{\omega_q}{V}\psi_M^\dagger(s)s \psi_M(s),
\end{eqnarray}
giving
\begin{eqnarray}
\Lambda_q &=& \frac{r}{\mathcal{L}}\frac{\lambda \omega_q}{V}\{\vec{\sigma}\cdot\hat{r}(2d\sinh (2\lambda r)-\cosh (2\lambda r)(\hat{r}\cdot\vec{d})+2(\hat{r}\cdot\vec{d}) \nonumber \\
&&+\mathcal{D}\frac{\sinh (2\lambda r)}{2\lambda r^2}\left(2\lambda r\coth (2\lambda r)-1\right))) -2\vec{\sigma}\cdot\vec{d} \}.
\label{eqn:LQ*}
\end{eqnarray}

The asymptotic expansion of $\Lambda^q$ is 
\begin{equation}
\Lambda_q=\vec{\sigma}\cdot\hat{r}\left(\frac{\lambda d\omega_q}{V}\frac{1}{r}-\frac{d\omega}{2}(d+\hat{r}\cdot\vec{d})\frac{1}{r^2}+O(r^{-3})\right).
\label{eqn:LQ*Asymp}
\end{equation}
From the asymptotic expansion we can see that there is no change to the framing of the monopole at infinity.

Now we must construct $iv^\dagger \delta_q v$. In Appendix B we have constructed many of the intermediate terms we use in calculating $iv^\dagger \delta_q v$.

We find that
\begin{eqnarray}
i v^\dagger \delta_q v &=& -\frac{r}{\mathcal{L}} \{\vec{\sigma}\cdot\hat{r} (-2 \sinh(2\lambda r) \omega_q d-\cosh(2\lambda r) \hat{r}\cdot\vec{R}_q d \nonumber\\
&&+\hat{r}\cdot\vec{R}_q d - \frac{N}{\mathcal{D}}\left( 4z\omega_q d-2\vec{z}\cdot\vec{R}_q d\right) ) \nonumber\\
&&-(\vec{\sigma}\times\hat{r})_q\left(\frac{a}{r} - \frac{a}{r} \cosh(2\lambda r) - \sinh(2\lambda r)+\frac{\mathcal{D}}{r}\left(\lambda-\frac{\sinh(2\lambda r)}{2r}\right)\right)\nonumber\\
&& -\vec{\sigma}\cdot\vec{R}_q d\},
\end{eqnarray}
where $\varphi^\dagger \delta_q\varphi=\hat{d}_q+i d \vec{\sigma}\cdot\vec{R}_q$.
The asymptotic expansion of $i v^\dagger \delta_q v$ is
\begin{eqnarray}
i v^\dagger \delta_q v &=& \vec{\sigma}\cdot\hat{r}\left( -\omega_q+\left( \omega_q d-\frac{\omega_q d}{2}\hat{r}\cdot\vec{d}\right)\frac{1}{r}+O(r^{-2})\right)\nonumber\\
&&+(\vec{\sigma}\times\hat{r})_q\left(\frac{1}{2r}+O(r^{-2})\right).
\label{eqn:vdvAsymp}
\end{eqnarray}

The combined asymptotic expansion of $\Omega_q=\Lambda_q-iv^\dagger \delta_q v$ is
\begin{eqnarray}
\Omega_q &=& \vec{\sigma}\cdot\hat{r}\left(\omega_q -\left( \frac{\omega_q}{2V}-\hat{r}\cdot\vec{d}\omega_q \frac{1/2+2d\lambda}{2V} \right) \frac{1}{r}+O(r^{-2})\right)\nonumber\\
&&+(\vec{\sigma}\times\hat{r})_q\left(\frac{1}{2r}+O(r^{-2})\right).
\label{eqn:OAsymp}
\end{eqnarray}

\section{Zero Modes of $SU(2)$ singular monopole}

Recall that our zero modes were defined as $Z^\mu_m=\delta_m A^\mu - D_\mu \Omega_m$.

Our gauge zero mode is
\begin{equation}
Z^\mu_0 = D_\mu \Lambda_0,
\label{eqn:Z0}
\end{equation}
with $\Lambda^0$ as computed in eq. (\ref{eqn:L0}).
The components of our translational zero modes are
\begin{eqnarray}
Z^0_q &=& \vec{\sigma}\cdot\hat{r}\left\{ \frac{r_q}{z}+2(\hat{z}_q+\hat{d}_q)\left(\lambda+\frac{1}{2z}\right)r\frac{\mathcal{D}}{\mathcal{L}}-\frac{\hat{z}_q}{z^2}\mathcal{K}\right\}\\
&& +\frac{r}{z}\vec{\sigma}\cdot\vec{d}_\perp\left\{\frac{\hat{z}_q}{z^2}-\frac{2}{\mathcal{L}}(\hat{z}+\hat{d}_q)N\right\}-\frac{r}{z}\sigma_q\nonumber\\
&&+\frac{\omega_q \lambda}{\mathcal{L}V}\frac{2r}{z\mathcal{L}}\{-\frac{2z}{r}\left(\left(\lambda+\frac{1}{2z}\right)\mathcal{K}-\frac{\mathcal{L}}{2r}\right)+2dS\nonumber\\
&&-2(\hat{r}\cdot\vec{d})C+\mathcal{D}\frac{S}{2\lambda r^2}\left(2\lambda r\frac{C}{S}-1\right))\vec{\sigma}\cdot(\vec{r}\times\vec{d})\}-D_0(iv^\dagger \delta_q v),\nonumber
\label{eqn:ZQ0}
\end{eqnarray}

\begin{eqnarray}
Z^i_q &=& (\vec{\sigma}\times\vec{r})_i\left\{(\hat{z}_q+\hat{d}_q)\left[\left(2\lambda+\frac{2a}{\mathcal{D}}\right)\left(a-\frac{N}{\mathcal{L}}\right)-\frac{2r}{z\mathcal{L}}N-\frac{a^2+r^2}{\mathcal{D}}\right]+r\frac{\hat{z}_q}{z^2}\right\}\nonumber\\
&&+(\vec{\sigma}\times\vec{d})_i\left\{r\frac{\hat{z}_q}{z^2}-\frac{2r}{z\mathcal{L}}N(\hat{z}_q+\hat{d}_q)\right\}\nonumber\\
&&-\vec{\sigma}\cdot\hat{r}\{\left(\frac{\mathcal{K}}{\mathcal{D}}-1\right)\left((\hat{r}\times\hat{z})_i\left(\frac{2r(\hat{z}_q+\hat{d}_q)}{\mathcal{L}}N-\frac{\hat{z}_q}{z^2}\right)+\frac{r}{z}(\hat{r}\times\hat{e}_q)_i\right)\nonumber\\
&&+\frac{2r(\hat{z}_q+\hat{d}_q)}{\mathcal{D}}(\hat{r}\times\hat{z})_i(M-a\frac{\mathcal{K}}{\mathcal{D}})\nonumber\\
&&+\frac{\omega_q \lambda}{\mathcal{L}V} \{ \left[2Sd-4C\hat{r}\cdot\vec{d}+4\hat{r}\cdot\vec{d}+\frac{\mathcal{D}S}{2\lambda r^2}\left(2\lambda r \frac{C}{S}-1\right)\right]\nonumber\\
&&\times\left[-\frac{2}{\mathcal{L}}(S(\hat{z}_i a+r_i)+C(\hat{r}_i a+r\hat{z}_i)+2\lambda \hat{r}_i \mathcal{K})-2\hat{r}_i\left(\left(\lambda+\frac{a}{\mathcal{D}}\right)\frac{\mathcal{D}}{\mathcal{L}}-\frac{1}{2r}-\frac{r^2}{z\mathcal{L}}\right)\right]\nonumber\\
&&+2C(2\lambda \hat{r}_i)(d-2S\hat{r}\cdot\vec{d})+2\hat{r}_i C+\left(2\lambda r\frac{C}{S}-1\right)\nonumber\\
&&\times\left[2S(\hat{z}_i a-r_i)\frac{1}{2\lambda r^2}+\frac{\mathcal{D}\hat{r}_i}{r^2}C-\frac{2\mathcal{D}\hat{r}_i}{r^3}S\right]+\frac{\mathcal{D}\hat{r}_i}{r^2}\left(C-\frac{2\lambda r}{S}\right)\}\}\nonumber\\
&&-\frac{r}{z}(\vec{\sigma}\times\hat{e}_q)_i+\vec{\sigma}\cdot\vec{d}\{\frac{2r}{z\mathcal{L}}\left(2Sd-4C\hat{r}\cdot\vec{d}+4\hat{r}\cdot\vec{d}+\frac{\mathcal{D}S}{2\lambda r^2}\left(2\lambda r\frac{C}{S}-1\right)\right)\nonumber\\
&&\times\left(2r\left(\left(\lambda+\frac{a}{\mathcal{D}}\right)\frac{\mathcal{D}}{\mathcal{L}}-\frac{r^2}{z\mathcal{L}}\right)-2\vec{r}\cdot\vec{d}\frac{r}{z\mathcal{L}}\right)\nonumber\\
&&+\frac{4d^2r^2}{z\mathcal{L}}-4\vec{r}\cdot\vec{d}\left(\left(\lambda+\frac{a}{\mathcal{D}}\right)\frac{\mathcal{D}}{\mathcal{L}}-\frac{1}{2r}-\frac{r^2}{z\mathcal{L}}\right)\}-D_i(iv^\dagger \delta_q v).
\label{eqn:ZQi}
\end{eqnarray}
These zero modes were independently verified as satisfying the linearized Bogomolny equations and the background gauge condition using Mathematica.

We find the asymptotic expansion of these zero modes as we will need these for the computation of the monopole moduli space metric in the next chapter.
\begin{eqnarray}
Z^0_0 &=& 0,\\
Z^i_0 &=& \vec{\sigma}\cdot\hat{r}(-\frac{1}{2V}\frac{r_i}{r^3}+O(r^{-4})),\\
Z_q^i&=& \vec{\sigma}\cdot\hat{r}\{\frac{(\hat{z}_q+\hat{d}_q)(\hat{r}\times\hat{z})_i}{2d^2(1+\hat{r}\cdot\hat{d})^2}\nonumber\\
&&+\frac{1}{r}\left(\frac{1}{d(1+\hat{r}\cdot\hat{d})}\right)\left((\hat{z}_q+\hat{d}_q)+\frac{(\hat{r}\times\hat{e}_q)_i}{2}-\frac{1-\hat{r}\cdot\hat{d}}{2(1+\hat{r}\cdot\hat{d})}\right)\nonumber\\
&& +\frac{1}{r^2}(-(\hat{z}_q+\hat{d}_q)(\hat{r}\times\hat{z})_i\left(\frac{7+6\hat{r}\cdot\hat{d}+(\hat{r}\cdot\hat{d})^2}{4(1+\hat{r}\cdot\hat{d})^2}\right)\nonumber\\
&&\frac{\omega_q}{2V}\hat{r}^i-\frac{\omega}{2V}(1+4\lambda d)(\hat{r}\cdot\hat{d})\hat{r}^i-\frac{(\hat{r}\times\hat{e}_q)_i}{2})\}.
\label{eqn:ZqAsymp}
\end{eqnarray}



\chapter{Metric of the Monopole Moduli Space} 
\label{Chapter3}
\lhead{Chapter 3. \emph{Metric of the Monopole Moduli Space}} 
We now use the zero modes found in the previous chapter to compute the metric components $g_{00}$ and $g_{0q}$ of the $SU(2)$ singular monopole's moduli space. We anticipate, based on \cite{CherkisKapustin1998}, that the moduli space is Taub-NUT. The Taub-NUT space is a self-dual Einstein manifold first described by Newman, Unti, and Tamubrino in \cite{NUT1963}.

\section{$g_{00}$}
We start with the relatively simple computation of $g_{00}$.
To calculate the metric component $g_{00}$ all we need is the gauge zero mode from eq. (\ref{eqn:Z0}) and the metric formula (\ref{eqn:metricformula}).
The metric component integral is
\begin{eqnarray}
g_{00} 	&=& \int d^3x Tr(Z^\mu_0Z^\mu_0),\nonumber\\
		&=& \int d^3x Tr(D_\mu \Lambda_0 D_\mu \Lambda_0),\nonumber\\
			&=& \int d^3x \frac{1}{2}Tr(D_\mu D_\mu(\Lambda_0^2)),\nonumber\\
			&=& \frac{1}{2}\int d^3x \partial_i\partial_iTr(\Lambda_0^2),\nonumber \\
			&=& \frac{1}{2}\int d^2S^i_\infty \partial_i Tr(\Lambda_0^2) -\int d^2S^i_{z\to 0}\partial_i Tr(\Lambda_0^2), 
\end{eqnarray}
where to go from the second to the third line we made use of the fact that $D_\mu D_\mu \Lambda_0=0$.
All partial derivatives $\partial_i$ are with respect to $r^i$.

Now we refer to $\Lambda_0$ and its asymptotic expansion as we calculated in the previous sections and in equations (\ref{eqn:L0})(\ref{eqn:L0Asymp}). We also have, from our definition of $\Lambda_0$ eq. (\ref{eqn:L0}), that $\Lambda_0$ is non-singular and smooth at the origin $z=0$ which means that the surface integral around the origin in the above equation vanishes.

The key ingredient in computing the metric component will be the $r^{-1}$ term in $Tr(\Lambda_0^2)$ so when we find the explicit form of $\Lambda_0$ we will be especially interested in the asymptotic expansion at $r\to\infty$ from eq. (\ref{eqn:L0Asymp}) .
\begin{equation}
\frac{1}{2}Tr(\Lambda_0^2) = 1-\frac{1}{V}\frac{1}{r}+O(r^{-2}),
 \end{equation}
 and the integral would be
 \begin{eqnarray}
 g_{00}=\frac{1}{2}\int d^2S^i_\infty \partial_i Tr(\Lambda_0^2) &=& \frac{1}{2}\int d\theta d\phi r^2 \sin(\theta) \hat{r}^i \partial_i Tr(\Lambda_0^2),\nonumber\\
 &=& \int d\theta d\phi r^2 \sin(\theta) \hat{r}^i \partial_i(1-\frac{1}{V}\frac{1}{r}+O(r^{-2})),\nonumber\\
 &=& \int d\theta d\phi r^2 \sin(\theta) \hat{r}^i \frac{1}{V} \frac{r_i}{r^3},\nonumber\\
 &=& \int d\theta d\phi \sin(\theta) \frac{1}{V}.
 \label{eqn:g00}
 \end{eqnarray}
 
 Thus our metric component $g_{00}$ is
 \begin{equation}
 g_{00}=\frac{4\pi}{V}.
 \end{equation}
 
 \section{$g_{0q}$}
 We compute
 \begin{equation}
 g_{0q} = \int d^3 x Tr(D_\mu \Lambda^0 Z^q_\mu),
 \end{equation}
 using our translational zero mode in equations (\ref{eqn:ZQi}).
 
 We take advantage of the fact that $D_\mu Z_q^\mu=0$ to write the term inside the integral as $D_\mu \Lambda^0 Z^q_\mu = D_\mu (\Lambda^0 Z^q_\mu)$. Then we take the trace of this term $Tr(D_\mu (\Lambda_0 Z_q^\mu))= \partial_\mu Tr(\Lambda_0 Z_q^\mu)=\partial_i Tr(\Lambda_0 Z_q^i)$ where we have also used the fact that the zero modes are static.
 This allows us to write $g_{0q}$ as a surface integral
 \begin{eqnarray}
 g_{0q} &=& \int d^2 S^i_\infty Tr(\Lambda_0 Z_q^i),\nonumber\\
 &=& \int d\theta d\phi r^2 \sin (\theta) Tr(\Lambda_0 \hat{r}^i Z_q^i),\nonumber\\
 &=& \int d\theta d\phi r^2 \sin(\theta) Tr(\Lambda_0 \hat{r}^i(\delta_q A^i-D_i\Omega_q)).
 \end{eqnarray}

Asymptotically we have $\Lambda_0 = (1-\frac{1}{2V}\frac{1}{r}+O(r^{-3}))\vec{\sigma}\cdot\hat{r}$ from eq.(\ref{eqn:L0Asymp}). We also have the asymptotic result
\begin{equation}
\hat{r}^i Z_q^i = \vec{\sigma}\cdot\hat{r}\left( \frac{\omega_q}{2V}-\frac{\omega_q}{2V}(1+4\lambda d)(\hat{r}\cdot\vec{d})\frac{1}{r^2}+O(r^{-3})\right),
\end{equation}
from our previous result eq. (\ref{eqn:ZqAsymp}).

With the asymptotic expansions it is simple to calculate the surface integral 
\begin{eqnarray}
\int d^2S^i_\infty Tr(\Lambda_0Z^q_i) &=& \int  d\theta d\phi r^2 \sin(\theta) Tr( \vec{\sigma}\cdot\hat{r}(1+O(r^{-1}) )\nonumber\\
&&\times \vec{\sigma}\cdot\hat{r}\frac{\omega_q}{2V}\left(\left(1-(1+4\lambda d)(\hat{r}\cdot\vec{d}) \right) \frac{1}{r^2}+O(r^{-3}) \right)),\nonumber\\
&=& \int  d\theta d\phi r^2 \sin (\theta) 2\left( \frac{\omega_q}{2V}\left(1-(1+4\lambda d)(\hat{r}\cdot\vec{d}) \right) \frac{1}{r^2}+O(r^{-3}) \right),\nonumber\\
&=& \int  d\theta d\phi \sin (\theta) \left( \frac{\omega_q}{V}\left(1-(1+4\lambda d)(\hat{r}\cdot\vec{d}) \right) \right), \nonumber\\
&=&4\pi \frac{\omega_q}{V},
\label{eqn:g0q}
\end{eqnarray}
where we used that $r\to\infty$ and $\int d\theta sin(\theta) \hat{r}\cdot\vec{d}=0$.

Thus our total metric component $g_{0q}$ is
\begin{equation}
g_{0q}=4\pi \frac{\omega_q}{V}.
\end{equation}

 \section{$g_{pq}$}
 The moduli space of our monopole is hyperk\"{a}hler \cite{CherkisKapustin1998} \cite{AtiyahHitchin1988} and has a triholonomic isometry. It suffices to know the metric components $g_{00}$ and $g_{0q}$ to completely fix the metric given these properties. Nevertheless, here we will now try to compute $g_{pq}$. We warn the reader that this is not complete and only gives a partial answer.
 
 \begin{equation}
g_{pq} = \int d^3x Tr(Z^\mu_p Z^\mu_q).
\label{eqn:gpq1}
 \end{equation}
 We can simplify the computation of this integral by extracting some boundary surface integrals.
 We recall that $Z_q^\mu=\delta_q A^\mu + D_\mu \Omega_q$.
So we have 
\begin{eqnarray}
Z_p^\mu Z_q^\mu &=& \delta_pA^\mu Z_q^\mu + D_\mu \Omega_p Z_q^\mu,\nonumber\\
&=& \delta_pA^\mu Z_q^\mu+D_\mu (\Omega_p Z_q^\mu),
\end{eqnarray}
where me made use of the condition $D_\mu Z_q^\mu=0$.
Now we turn our attention to the first term on the second line, $\delta_pA^\mu Z_q^\mu$,
\begin{equation}
\delta_pA^\mu Z_q^\mu = \delta_pA^\mu \delta_qA^\mu+\delta_pA^\mu D_\mu\Omega_q.
\end{equation}
We can rewrite the second term on the previous line as
\begin{equation}
\delta_pA^\mu D_\mu\Omega_q = D_\mu (\delta_p A^\mu \Omega_q)-D_\mu \delta_p A^\mu \Omega_q.
\label{eqn:gpqinter}
\end{equation}
We make use of the identity $D_\mu Z_p^\mu = D_\mu \delta_p A^\mu + D_\mu D_\mu \Omega_p=0$. So $D_\mu \delta_p A^\mu \Omega_q = -D_\mu D_\mu \Omega_q \Omega_p$.
We can rewrite the last term in equation (\ref{eqn:gpqinter})
\begin{equation}
D_\mu D_\mu \Omega_p \Omega_q = D_\mu (D_\mu \Omega_p \Omega_q)-D_\mu \Omega_q D_\mu \Omega_p.
\end{equation}

We use this result to separate the equation (\ref{eqn:gpq1}) into a surface boundary integral and a volume integral
\begin{eqnarray}
g_{pq} &=& \int d^3x Tr(Z_p^\mu Z_q^\mu),\nonumber\\
&=&\int d^2S^i_\infty Tr(\Omega_p Z_q^i+Z_p^i\Omega_q),\nonumber\\
&&+\int d^3x Tr(\delta_pA^\mu\delta_qA^\mu-D_\mu\Omega_pD_\mu\Omega_q).\nonumber\\
\label{eqn:gpq2}
\end{eqnarray}

We can use our previously found asymptotic results for $\Omega^q$ eq. (\ref{eqn:OAsymp}) and $Z^q_i$ eq. (\ref{eqn:ZqAsymp}) to compute the surface integrals 
\begin{equation}
\int d^2S^i_\infty Tr(\Omega^q Z^p_i) = \int d^2S^i_\infty Tr(Z^q_i\Omega^p)=4\pi\frac{\omega_q\omega_p}{2V}.
\label{eqn:gpqsurf}
\end{equation}

Plugging this into equation (\ref{eqn:gpq2}) we get
\begin{eqnarray}
g_{pq} &=& \int d^3x Tr(Z_p^\mu Z_q^\mu),\nonumber\\
&=&4\pi\frac{\omega_q\omega_p}{V}+\int d^3x Tr(\delta_pA^\mu\delta_qA^\mu-D_\mu\Omega_pD_\mu\Omega_q).\nonumber\\
\label{eqn:gpq}
\end{eqnarray}

We anticipate, but have not computed, that the volume integral will be 
\begin{equation}
\int d^3x Tr(\delta_pA^\mu\delta_qA^\mu-D_\mu\Omega_pD_\mu\Omega_q)=4\pi V \delta_{pq},
\end{equation} where $\delta_{pq}$ is the Kronecker delta. We will calculate the moduli space metric over the Nahm data in the next chapter and in chapter 5 we will connect this result with the monopole moduli space.
In conclusion the only hyperk\"{a}hler metric with $U(1)$ isometry and $g_{00}=4\pi\frac{1}{V}$ and $g_{0q}=4\pi\frac{\omega_q}{V}$ as we computed in sections 3.1 and 3.2 is

\begin{equation}
ds^2=4\pi\left( Vd\vec{T}^2+\frac{1}{V}\left(dT_0+\vec{\omega}\cdot d\vec{T}\right)^2\right).
\end{equation}


\chapter{Metric of the Moduli Space of Nahm Data} 
\label{Chapter4}
\lhead{Chapter 4. \emph{Metric of the Nahm Moduli Space}} 

It has been shown by Nakajima \cite{Nakajima1990} and Maciocia \cite{Maciocia} that the metric of the Nahm moduli space and the metric of the monopole moduli space produced by the Nahm transform are isomorphic for regular monopoles. We calculate the metric of the Nahm moduli space in this chapter. We use the hyperk\"{a}hler quotient construction method, as described in \cite{GibbonsEtAl} \cite{CherkisKapustin1999}.

We perform the hyperk\"{a}hler quotient construction of the metric over the Nahm data by first finding the metric over $T^\mu$ with a trivial gauge action at the boundaries and the metric over the jumping data $\varphi$ with non-trivial gauge action, and then we take the direct product of these spaces and mod out the $U(1)$ gauge action finding the metric on the moduli space as a $U(1)$ hyperk\"{a}hler quotient.

Let us review our Nahm data for our $SU(2)$ singular monopole construction
\begin{equation}
\vec{T}(s) = \vec{T}\Theta'(s),
\end{equation}  
where $\vec{T}=\vec{d}$ and $\Theta'(s)=\Theta(s+\lambda)-\Theta(s-\lambda)$. We also have the jumping data spinors $\varphi=\begin{pmatrix}f^\dagger_+\\f^\dagger_-\end{pmatrix}$. 

The tangent vector over the Nahm data $T^\mu$ is $(dt^0,dT^1,dT^2,dT^3)$ where $dt_0$ is the collective coordinate associated with the gauge action. We consider a group action that is trivial at the endpoints of the interval $(-\lambda,\lambda)$ and non-trivial in between the endpoints: $g(s)=e^{if(s)}$ where $f(\pm\lambda)=2\pi n$, $n\in\mathbb{Z}$. The group action is $t^0\to t^0-\partial_s f(s)$.  Due to the condition of triviality of the group action at the end points $g(s)=e^{i\frac{s+\lambda}{\lambda}\pi n}: t^0\to t^0+\frac{\pi}{\lambda}n$ , we determine that $t^0\in S^1$, a periodic coordinate. 
Therefore the metric on just the Nahm data with trivial gauge action at the endpoints is

\begin{equation}
ds^2 = \int^\lambda_{-\lambda} ds Tr\left( (dt^0)^2I_{2\times 2}+(d\vec{T})^2I_{2\times 2}\right) = 4\lambda \left( (dt^0)^2+(d\vec{T})^2\right).
\end{equation} 

The moduli space is $S^1 \times \mathbb{R}^3$, where $S^1$ is of radius $\frac{1}{\sqrt{\lambda}}$.

That leaves us to consider the metric from a group action which is non-trivial at $s=\pm\lambda$. Such a non-trivial group action $g(s)$ acts on the jumping data as $f_+\to g(\lambda)f_+$ and $f_-\to g(-\lambda)f_-$.

The metric over the jumping data is \cite{GibbonsEtAl}
\begin{equation}
2Tr\left(\delta\varphi^\dagger\delta\varphi\right)=\frac{1}{2d}(d\vec{r})^2+2d\left( d\psi+\vec{\omega}\cdot d\vec{r}\right)^2,
\end{equation}
where in the above equation $\vec{\sigma}\cdot\vec{r} = \varphi^\dagger \sigma_3\varphi$, $r=|\vec{r}|$, and $\varphi = e^{i\sigma_3 \psi/2} a$ where $a$ is pure imaginary $a^\dagger=-a$. 

As is shown in \cite{GibbonsEtAl} as well as \cite{Cherkis2009} the moment maps of the gauge action tell us that $\frac{1}{2}\vec{r}=\vec{T}$. We combine these two metrics to get
\begin{eqnarray}
ds^2 &=& 4\lambda d\vec{T}^2+4\lambda (dt_0)^2 + \frac{1}{2d} d\vec{r}^2+2d\left( d\psi+\vec{\omega}\cdot d\vec{r}\right)^2,\nonumber\\
&=& 4\left( \left(\lambda+\frac{1}{2d}\right) d\vec{T}^2 +\lambda (dt_0)^2+\frac{d}{2}\left(d\psi+2\vec{\omega}\cdot d\vec{T}\right)^2\right),\nonumber\\
&=&4\left( V d\vec{T}^2 +\lambda (dt_0)^2+\frac{d}{2}\left(d\psi+2\vec{\omega}\cdot d\vec{T}\right)^2\right).\nonumber\\
\label{eqn:combmetric}
\end{eqnarray}
after we use the moment map relation we described above and $V=\lambda+\frac{1}{2d}$.

We now define a quantity $T_0$ that is invariant under non-trivial $U(1)$ gauge action at the boundaries: $2T_0 = \psi+2\lambda t_0$ and find $d\psi=2(dT_0)-2\lambda (dt_0)$. We substitute this back into $\lambda (dt_0)^2+\frac{d}{2}\left( 2(dT_0)+2\vec{\omega}\cdot d\vec{T}-2\lambda (dt_0)\right)^2$
\begin{eqnarray}
&=& \lambda(dt_0)^2+2d\left((dT_0)+\vec{\omega}\cdot d\vec{t}-\lambda (dt_0)\right)^2,\nonumber\\
&=& \left(\lambda+2\lambda^2 d\right)(dt_0)^2-4\lambda d (dt_0)\left( dT_0+\vec{\omega}\cdot d\vec{T}\right)+2d\left( dT_0+\vec{\omega}\cdot d\vec{T}\right)^2,\nonumber\\
&=&\left( \lambda+2\lambda^2 d\right) \left( dt_0-\frac{2\lambda d}{\lambda+2\lambda^2 d}\left( dT_0+\vec{\omega}\cdot d\vec{T}\right) \right)^2,\nonumber\\
&&+\left( 2d-\frac{2\lambda d}{\lambda+2\lambda^2 d}\right)\left( dT_0+\vec{\omega}\cdot d\vec{T}\right)^2,\nonumber
\end{eqnarray}
where we have expanded the terms and completed the square. The first term on the last line is the only quantitiy acted on by the group action and is therefore modded out. The second term on the last line is simplified to $\frac{1}{V}\left((dT_0)+\vec{\omega}\cdot d\vec{T}\right)^2$.

We plug this in back into eq.(\ref{eqn:combmetric}) and get the metric on the Nahm moduli space
\begin{equation}
ds^2 = 4\left(V (d\vec{T})^2+\frac{1}{V}\left(dT_0+\vec{\omega}\cdot d\vec{T}\right)^2\right).
\label{eqn:hkq}
\end{equation}

Equivalently, rather than quotient out the group action from the combined metric, we can also find tangent vectors which are orthogonal to small gauge transformations and calculate their overlap to construct the metric. This method is similar to the direct computation of the moduli space metric discussed in \cite{Kraan}, and is also similar to the way we constructed the monopole moduli space metric in Chapter 3.
We adopt the notation $<h>_N=\int^\infty_{-\infty} ds h(s)$. 
The Nahm moduli space metric is defined as
\begin{equation}
ds^2=\left<Tr(H^\dagger H)\right>_N=Tr\left(\left<\hat{Y}^\dagger\hat{Y}\right>_N+2\left<\hat{c}^\dagger\right>_N\left<\hat{c}\right>_N\right).
\end{equation}
We have, as in (\ref{eqn:HDef}), our Nahm tangent vector $H=\begin{pmatrix}\hat{c}\\ \hat{Y}\end{pmatrix}$ which satisfies the Nahm background gauge condition, and where $\hat{Y}$ and $\hat{c}$ are, as a reminder, defined as follows  $\hat{Y}=\Theta' (i\vec{\sigma}\cdot d\vec{T}+\frac{1}{V}(dT^0+\vec{\omega}\cdot d\vec{T})I_{2\times 2})$ and $\hat{c}=\delta(s+\lambda)P_2(-i\delta\varphi-\frac{\lambda}{V}(dT^0+\vec{\omega}\cdot d\vec{T})\varphi)+\delta(s-\lambda)P_1(-i\delta\varphi+\frac{\lambda}{V}(dT^0+\vec{\omega}\cdot d\vec{T})\varphi)$ where we have chosen that $\chi^q=0$ as in section 2.1.1. 

We calculate in a straightforward manner 
\begin{equation}
Tr\left<\hat{Y}^\dagger\hat{Y}\right>_N=4\lambda (d\vec{T})^2+\frac{4\lambda}{V^2}(dT^0+\vec{\omega}\cdot d\vec{T})^2,
\label{eqn:YY}
\end{equation}
\begin{eqnarray}
\left<\hat{c}\right>_N&=&-i\delta\varphi+\frac{\lambda}{V}(dT^0+\vec{\omega}\cdot d\vec{T})\sigma_3 \varphi,\\
\left< \hat{c}^\dagger \right>_N&=&i\delta\varphi^\dagger+\frac{\lambda}{V}(dT^0+\vec{\omega}\cdot d\vec{T})\varphi^\dagger\sigma_3.
\end{eqnarray}
So using these results we calculate
\begin{equation}
2Tr\left<\hat{c}^\dagger\right>_N\left<\hat{c}\right>_N =\frac{2}{d}(d\vec{T})^2+8d\left(1-\frac{\lambda}{V}+\frac{\lambda^2}{V^2}\right)(dT^0+\vec{\omega}\cdot d\vec{T})^2.
\label{eqn:cc}
\end{equation}

We combine equations (\ref{eqn:YY}) and (\ref{eqn:cc}) and find
\begin{equation}
ds^2 = 4\left(V (d\vec{T})^2+\frac{1}{V}(dT^0+\vec{\omega}\cdot d\vec{T})^2\right),
\label{eqn:directnahm}
\end{equation}
which matches eq. (\ref{eqn:hkq}) and is the Taub-NUT metric.

In the previous chapter we compute the metric on the moduli space of the singular monopole. In this chapter we computed the metric on the associated Nahm data. Therefore we have explicitly verified the isometry between these two moduli spaces.


\chapter{Corrigan's Formula} 
\label{Chapter5}
\lhead{Chapter 5. \emph{Corrigan's Formula}} 

We can also take advantage of the Nahm construction and Corrigan's inner product formula to prove the isometry of the metrics we computed in Chapters 3 and 4.

We recast $Tr(Z_m^\mu Z_n^\mu)$ as a second order derivative using a remarkably useful formula first devised by Corrigan in an unpublished work that was quoted and used by Osborn in  \cite{Osborn1981}. 

Using our previous definitions of $Z_m^\mu$ in eq. (\ref{eqn:ZDef}) and $H$ in eq. (\ref{eqn:HDef}) we can rewrite $Tr(Z_m^\mu Z_n^\mu)$ per Corrigan's formula (as applied by \cite{KraanVanBaal1998}):
\begin{eqnarray}
Tr(Z_m^\mu Z_n^\mu) &=& -\frac{1}{2} \partial^2 Tr \int^\infty_{-\infty} ds F(s,s)(\hat{Y}^\dagger_m(s)\hat{Y}_n(s)+\hat{Y}^\dagger_n(s)\hat{Y}_m(s)\nonumber\\
&& +\hat{c}^\dagger_m(s)<\hat{c}_n>+\hat{c}^\dagger_n<\hat{c}_m>)\nonumber\\
&&+\frac{1}{2}\partial^2 Tr\int^\infty_{-\infty} ds dt F(t,s)([\hat{c}_m\hat{\varphi}+\hat{Y}^\dagger(s)_m\hat{D}(s)]\nonumber\\
&&F(s,t)[\hat{D}^\dagger(t)\hat{Y}_n(t)+\hat{\varphi}^\dagger\hat{c}_n]),
\label{eqn:Corrigan}
\end{eqnarray}
where from our construction of the zero modes we have $\hat{Y}=\Theta' (i\vec{\sigma}\cdot d\vec{T}+\frac{1}{V}(dT^0+\vec{\omega}\cdot d\vec{T})I_{2\times 2})$ and $\hat{c}=\delta(s+\lambda)P_2(-i\delta\varphi-\frac{\lambda}{V}(dT^0+\vec{\omega}\cdot d\vec{T})\varphi)+\delta(s-\lambda)P_1(-i\delta\varphi+\frac{\lambda}{V}(dT^0+\vec{\omega}\cdot d\vec{T})\varphi)$.
An explicit derivation of the preceding formula can be found in \cite{Dorey1996}. 
We adapt the use of this formula from \cite{Kraan} \cite{KraanVanBaal1998} to our singular monopole case. 

We integrate the volume integral for the metric component $g_{mn}$ by parts to get a surface integral
\begin{eqnarray}
g_{mn}  = \int d^3x Tr(Z_m^\mu Z_n^\mu) &=&  -\frac{1}{2} \int d^2S^i_\infty \partial_i Tr \int^\infty_{-\infty} ds F(s,s)(\hat{Y}^\dagger_m(s)\hat{Y}_n(s)+\hat{Y}^\dagger_n(s)\hat{Y}_m(s)\nonumber\\
&& +\hat{c}^\dagger_m(s)<\hat{c}_n>+\hat{c}^\dagger_n<\hat{c}_m>)\nonumber\\
&&+\frac{1}{2}\int d^2S^i_\infty \partial_i Tr\int^\infty_{-\infty} ds \int^\infty_{-\infty} dt F(t,s)([\hat{c}_m\hat{\varphi}+\hat{Y}_m^\dagger(s)\hat{D}(s)]\nonumber\\
&&F(s,t)[\hat{D}^\dagger(t)\hat{Y}_n(t)+\hat{\varphi}^\dagger\hat{c}_n]).
\label{eqn:CorriganSurf}
\end{eqnarray}

To simplify the above formula we note that asymptotically the covariant Laplacian becomes $\hat{D}^\dagger(s) \hat{D}(s)=-\partial^2_s+r^2$, so the asymptotic Green's function is defined by the Sturm-Liouville equation 
\begin{equation}
\hat{D}^\dagger(s)\hat{D}(s)F(s,t)=-\partial^2_s F(s,t)+r^2 F(s,t)=\delta(s-t).
\end{equation}

From this we determine that asymptotically \begin{equation}
F(s,t) = e^{r|s-t|}\left(-\frac{1}{2r}+O(r^{-2})\right).
\end{equation}
This means that $F(s,s)=-\frac{1}{2r}+O(r^{-2})$, $F(s,\pm\lambda)=e^{r|s\mp\lambda|}\left(-\frac{1}{2r}+O(r^{-2})\right)$, and $F(\pm\lambda,t)=e^{r|\pm\lambda-t|}\left(-\frac{1}{2r}+O(r^{-2})\right)$. We also have asymptotically that
\begin{equation}
\hat{D}(s)F(s,t)\hat{D}^\dagger(t)=\delta(s-t),
\end{equation}
and $F^2(s,s)=O(r^{-2})$.

Since we are investigating surface integrals in eq. (\ref{eqn:CorriganSurf}) we are only interested in the $O(r^{-1})$ part of each term in the integral. Let us investigate the terms separately in the surface integral in the 3rd and 4th lines of eq. (\ref{eqn:CorriganSurf}).

First we have 
\begin{eqnarray}
\int ds\int dt F(t,s) \hat{Y}_m^\dagger(s)\hat{D}(s)F(s,t)\hat{D}^\dagger(t)\hat{Y}_n(t),\nonumber\\
&=&\int ds\int dt F(t,s) \hat{Y}_m^\dagger(s)\delta(s-t)\hat{Y}_n(t), \nonumber\\
&=& \int ds \hat{Y}_m^\dagger(s)\hat{Y}_n(s) F(s,s),
\end{eqnarray} 
which we combine with the terms on the first line of eq. (\ref{eqn:CorriganSurf}). We also have the terms $\int ds\int dt F(t,s)\hat{c}^\dagger_m \hat{\varphi}F(s,t)\hat{\varphi}^\dagger\hat{c}_n = O(r^{-2})$ and  $\int ds \int dt \hat{Y}_m^\dagger(s)\hat{D}(s)F(s,t)\hat{\varphi}^\dagger\hat{c}_n= \int ds \hat{Y}_m^\dagger(s)\hat{D}(s)F(s,\lambda)\hat{\varphi}\hat{c}_n F(\lambda,s)+\int ds  \hat{Y}_m^\dagger(s)\hat{D}(s)F(s,-\lambda)\hat{\varphi}\hat{c}_n F(-\lambda,s)=O(r^{-2})$ after integration.  

We combine these results and the asymptotic behavior of $F$ into eq. (\ref{eqn:CorriganSurf}) to find the following expression for the metric
\begin{eqnarray}
\int d^3x Tr\left( Z_m^\mu Z_n^\mu\right) &=& \frac{1}{4}\int d\theta d\phi \sin(\theta) Tr\left( \left< \hat{Y}_m^\dagger \hat{Y}_n\right>_N + 2\left<\hat{c}^\dagger_m\right>_N\left<\hat{c}_n\right>_N\right),\nonumber\\
&=& \pi Tr\left(<\hat{Y}^\dagger_m\hat{Y}_n>_N+2\left<\hat{c}^\dagger_m\right>_N\left<\hat{c}_n\right>_N\right).
\label{eqn:ModMetric}
\end{eqnarray}
We note that $Tr\left(\left<\hat{Y}^\dagger\hat{Y}\right>_N+2\left<\hat{c}^\dagger\right>_N\left<\hat{c}\right>_N\right)$ is the Nahm metric which we have already calculated in the previous chapter.
\begin{equation}
Tr\left(\left<\hat{Y}^\dagger\hat{Y}\right>_N+2\left<\hat{c}^\dagger\right>_N\left<\hat{c}\right>_N\right)=4\left(V (d\vec{T})^2+\frac{1}{V}(dT^0+\vec{\omega}\cdot d\vec{T})^2\right).
\end{equation}

So after simply inserting the Nahm metric into eq. (\ref{eqn:ModMetric}) we find
\begin{equation}
ds^2=g_{mn}dT^mdT^n=4\pi\left(V d\vec{T}^2+\frac{1}{V}(dT^0+\vec{\omega}\cdot d\vec{T})^2\right),
\label{eqn:corriganmetric}
\end{equation} 
where $g_{00}=\frac{4\pi}{V}$ jand $g_{0q}=4\pi\frac{\omega_q}{V}$ just as we calculated in Chapter 3, equations (\ref{eqn:g00}) and (\ref{eqn:g0q}). 


\chapter{Conclusion} 
\label{Chapter6}
\lhead{Chapter 6. \emph{Conclusion}} 

We found the following results in this thesis. We explicitly constructed the phase zero mode and translational zero modes of the $SU(2)$ monopole with one singularity. These zero modes can be found in equations (\ref{eqn:Z0})(\ref{eqn:ZQi}) and their relevant asymptotic expansions in equations (\ref{eqn:L0Asymp})(\ref{eqn:LQ*Asymp})(\ref{eqn:vdvAsymp})(\ref{eqn:OAsymp})(\ref{eqn:ZqAsymp}).

We then computed some of the metric components of the moduli space of the monopole using these zero modes and their asymptotic expansions. The $g_{00}$ component is found in section 3.1 and in equation (\ref{eqn:g00}). The $g_{0q}$ component is in section 3.2 and in equation (\ref{eqn:g0q}). The $g_{pq}$ component calculation can be found in section 3.3 and is separated into a surface integral boundary term part that we evaluated in eq. (\ref{eqn:gpqsurf}) and a volume integral part in equation (\ref{eqn:gpq}). The surface boundary term is computed. Further work would include the computation of the volume integral in this equation.

To compare with our monopole moduli space, we computed the moduli space of the Nahm data, using the hyperk\"{a}hler quotient construction in eq. (\ref{eqn:hkq}) and direct computation using the Nahm zero modes in eq. (\ref{eqn:directnahm}).

Finally we directly connected the moduli space over the Nahm data with the $SU(2)$ singular monopole moduli space using Corrigan's inner product formula eq. (\ref{eqn:Corrigan}) and the asymptotic behavior of the monopole's Green's function. Thus we provided an independent proof of the isometry of the moduli space of the Nahm data and that of the singular monopole. 



\addtocontents{toc}{\vspace{2em}} 

\pagestyle{plain}
\appendix 


\chapter{Construction of $\Lambda^q$}

In order to go from eq. (\ref{eqn:NahmBGC}) to eq. (\ref{eqn:NahmBGC2}) we directly compute the trace terms in eq. (\ref{eqn:NahmBGC}).
\begin{eqnarray}
Tr\left( \varphi^\dagger P_{1,2} \varphi\right) &=& 2d\nonumber\\
-iTr\left(\varphi^\dagger P_1\delta \varphi-\delta\varphi^\dagger P_1\varphi\right) &=& 2d\left(dT^0+\vec{\omega}\cdot d\vec{T}\right)\nonumber\\
-iTr\left(\varphi^\dagger P_2\delta \varphi-\delta\varphi^\dagger P_2\varphi\right) &=& -2d\left(dT^0+\vec{\omega}\cdot d\vec{T}\right)
\label{eqn:trterms}
\end{eqnarray}. 

We describe here the intermediate terms in the construction of the singular monopole and also the gauge terms $\Lambda^0$ and $\Lambda^q$.
First we compute $\Delta^\dagger_{-}\Delta_{-}$ and $\Delta^\dagger_{+}\Delta_{+}$.
\begin{eqnarray}
\Delta^\dagger_{-}\Delta_{-} &=& N^2 \frac{r}{\sinh (2\lambda r)}\frac{1}{\mathcal{D}}e^{-\vec{\sigma}\cdot\vec{r}\lambda}\zeta_-\zeta^\dagger_-e^{-\vec{\sigma}\cdot\vec{r}\lambda}\nonumber\\
&=& \frac{r}{\mathcal{L}}(\cosh (\lambda r)-\vec{\sigma}\cdot\hat{r}\sinh(\lambda r))(z-\vec{\sigma}\cdot\vec{z})(\cosh (\lambda r)-\vec{\sigma}\cdot\hat{r}\sinh(\lambda r))\nonumber
\end{eqnarray}

\begin{eqnarray}
\Delta^\dagger_{+}\Delta_{+} &=& N^2 \frac{r}{\sinh (2\lambda r)}\frac{1}{\mathcal{D}}e^{\vec{\sigma}\cdot\vec{r}\lambda}\zeta_+\zeta^\dagger_+e^{\vec{\sigma}\cdot\vec{r}\lambda}\nonumber\\
&=& \frac{r}{\mathcal{L}}(\cosh (\lambda r)+\vec{\sigma}\cdot\hat{r}\sinh(\lambda r))(z+\vec{\sigma}\cdot\vec{z})(\cosh (\lambda r)+\vec{\sigma}\cdot\hat{r}\sinh(\lambda r))\nonumber
\end{eqnarray}

We then simplify $\Delta^\dagger_{+}\Delta_{+}-\Delta^\dagger_{-}\Delta_{-}$
\begin{equation}
\Delta^\dagger_{+}\Delta_{+}-\Delta^\dagger_{-}\Delta_{-} = \frac{r}{\mathcal{L}}\left(2z\sinh (2\lambda r)\vec{\sigma}\cdot\hat{r}+2\vec{\sigma}\cdot\vec{z}+(2\cosh (2\lambda r)-2)(\vec{z}\cdot\hat{r})\vec{\sigma}\cdot\hat{r}\right)
\label{eqn:DeltaInter}
\end{equation}

Next we'll compute $\int ds \psi_L^\dagger\psi_L$ and $\int ds \psi_R^\dagger\psi_R$.
\begin{eqnarray}
\int ds \psi^\dagger_L\psi_L &=& \int ds N^2 \frac{r}{\sinh (2\lambda r)} \frac{1}{\mathcal{D}} 2z e^{-\vec{\sigma}\cdot\vec{r}\lambda}(d+\vec{\sigma}\cdot\vec{d})e^{-\vec{\sigma}\cdot\vec{r}\lambda} e^{2z(s+\lambda)}\nonumber \\
&=& \int ds \frac{2rz}{\mathcal{L}}e^{-\vec{\sigma}\cdot\vec{r}\lambda}(d+\vec{\sigma}\cdot\vec{d})e^{-\vec{\sigma}\cdot\vec{r}\lambda}e^{2z(s+\lambda)}
\end{eqnarray}
\begin{eqnarray}
\int ds \psi^\dagger_R\psi_R &=& \int ds N^2 \frac{r}{\sinh (2\lambda r)} \frac{1}{\mathcal{D}} 2z e^{\vec{\sigma}\cdot\vec{r}\lambda}(d-\vec{\sigma}\cdot\vec{d})e^{\vec{\sigma}\cdot\vec{r}\lambda} e^{-2z(s-\lambda)}\nonumber \\
&=& \int ds \frac{2rz}{\mathcal{L}}e^{\vec{\sigma}\cdot\vec{r}\lambda}(d-\vec{\sigma}\cdot\vec{d})e^{\vec{\sigma}\cdot\vec{r}\lambda}e^{-2z(s-\lambda)}
\end{eqnarray}
The integrals  $\int^{-\lambda}_{-\infty} ds e^{2z(s+\lambda)} = \frac{1}{2z}$ and $\int^\infty_\lambda ds e^{-2z(s-\lambda)} = \frac{1}{2z}$.

We simplify $\int ds \left(\psi_R^\dagger\psi_R-\psi_L^\dagger\psi_L\right)$
\begin{equation}
\int ds \left(\psi_R^\dagger\psi_R-\psi_L^\dagger\psi_L\right) = \frac{r}{\mathcal{L}}(2d\sinh (2\lambda r) \vec{\sigma}\cdot\hat{r}-2\vec{\sigma}\cdot\vec{d}-(2\cosh (2\lambda r)-2)(\hat{r}\cdot\vec{d}) \vec{\sigma}\cdot\hat{r})
\label{eqn:PsiLRInter}
\end{equation}

Finally we compute $\int ds s\psi^\dagger_M\psi_M$

\begin{eqnarray}
\int ds s\psi^\dagger_M\psi_M &=&\int N^2 \frac{r}{\sinh (2\lambda r)} se^{2\vec{\sigma}\cdot\vec{r}s}\nonumber\\
&=& \int ds \frac{r\mathcal{D}}{\mathcal{L}} se^{2\vec{\sigma}\cdot\vec{r}s}\nonumber\\
&=& \frac{\mathcal{D}\sinh (2\lambda r)}{\mathcal{L}} \frac{1}{2r^2}\left( 2\lambda r\coth (2\lambda r)-1\right) \vec{\sigma}\cdot\hat{r}
\label{eqn:PsiMInter}
\end{eqnarray}

The Green's functions for the covariant Laplacian would be as follows and are cited directly from \cite{Durcan2007}
\begin{eqnarray}
F(s<-\lambda,t\in(-\lambda,\lambda))	&=&	\frac{e^{z(s+\lambda)}}{\mathcal{L}}(r\cosh r(t-\lambda)-a\sinh r(t-\lambda))\nonumber\\
F(s\in(-\lambda,\lambda),t\in(-\lambda,\lambda))	&=&	\frac{1}{2r\mathcal{L}}\{\cosh r(s+t)(r^2-a^2)\nonumber\\
+\cosh r|s-t|\mathcal{K}-\sinh r|s-t|\mathcal{L}\} \nonumber \\
F(s>\lambda,t\in(-\lambda,\lambda)) 	&=& 	\frac{e^{-z(s-\lambda)}}{\mathcal{L}}(r\cosh r(t+\lambda)+a\sinh r(t+\lambda))\nonumber\\
F(s<-\lambda,t>\lambda)	&=& 	\frac{re^{z(s-t+2\lambda)}}{\mathcal{L}}\nonumber\\
F(s\in(-\lambda,\lambda),t>\lambda)	&=&	\frac{e^{-z(t-\lambda)}}{\mathcal{L}}(r\cosh r(s+\lambda)+a\sinh r(s+\lambda))\nonumber\\
F(s>\lambda,t>\lambda) 	&=&	\frac{e^{-z|s-t|}}{2z}\nonumber\\
								&&		-e^{-z(s+t-2\lambda)}\frac{\mathcal{L}-2z(r\cosh 2\lambda r+a\sinh 2\lambda r)}{2z\mathcal{L}} \nonumber
\end{eqnarray}
Note that $F(s,t<\lambda)=F(-s,t>\lambda)$.


\chapter{Construction of $i v^\dagger \delta_q v$}

In this appendix we construct $iv^\dagger \delta_q v$ where $v$ is defined in eq. (\ref{eqn:PsiDef}) and $\delta_q=\frac{\partial}{\partial d_q}$.

First we find $\delta_q v$.
\begin{eqnarray}
\delta_q \Delta_+ &=& \Delta_+ \left( -\frac{\zeta^\dagger_+ \delta_q f_+}{\zeta^\dagger_+f_+} +\sqrt{\frac{\mathcal{L}}{\mathcal{D}r}} \delta_q\sqrt{\frac{\mathcal{D}r}{\mathcal{L}}}+e^{-\vec{\sigma}\cdot\vec{r}\lambda}\delta_qe^{\vec{\sigma}\cdot\vec{r}\lambda}\right)\nonumber\\
\delta_q \Delta_-&=&\Delta_- \left( \frac{\zeta^\dagger_- \delta_q f_-}{\zeta^\dagger_-f_-}+\sqrt{\frac{\mathcal{L}}{\mathcal{D}r}}\delta_q\sqrt{\frac{\mathcal{D}r}{\mathcal{L}}}+e^{\vec{\sigma}\cdot\vec{r}\lambda}\delta_qe^{-\vec{\sigma}\cdot\vec{r}\lambda}\right)\nonumber\\
\delta \psi_L &=& \psi_L\left(-\frac{\delta_qf^\dagger_+\zeta_+}{f^\dagger_+\zeta_+}+e^{\vec{\sigma}\cdot\vec{r}\lambda}\delta_qe^{-\vec{\sigma}\cdot\vec{r}\lambda}+\sqrt{\frac{\mathcal{L}}{\mathcal{D}r}}\delta_q\sqrt{\frac{\mathcal{D}r}{\mathcal{L}}}\right)\nonumber\\
&&+e^{z(s+\lambda)}\frac{\zeta_+\delta_qf^\dagger_+}{f^\dagger_+\zeta_+}e^{-\vec{\sigma}\cdot\vec{r}\lambda}\sqrt{\frac{\mathcal{D}r}{\mathcal{L}}}\nonumber\\
\delta \psi_R &=& \psi_R\left(-\frac{\delta_qf^\dagger_-\zeta_-}{f^\dagger_-\zeta_-}+e^{-\vec{\sigma}\cdot\vec{r}\lambda}\delta_qe^{\vec{\sigma}\cdot\vec{r}\lambda}+\sqrt{\frac{\mathcal{L}}{\mathcal{D}r}}\delta_q\sqrt{\frac{\mathcal{D}r}{\mathcal{L}}}\right)\nonumber\\
&&+e^{-z(s-\lambda)}\frac{\zeta_-\delta_qf^\dagger_-}{f^\dagger_-\zeta_-}e^{\vec{\sigma}\cdot\vec{r}\lambda}\sqrt{\frac{\mathcal{D}r}{\mathcal{L}}}\nonumber\\
\delta_q \psi_M &=& \psi_M\left( e^{-\vec{\sigma}\cdot\vec{r}s}\delta_qe^{\vec{\sigma}\cdot\vec{r}s}+\sqrt{\frac{\mathcal{L}}{\mathcal{D}r}}\delta_q\sqrt{\frac{\mathcal{D}r}{\mathcal{L}}}\right)
\end{eqnarray}

We then use these terms to construct $iv^\dagger \delta_q v$:
\begin{eqnarray}
iv^\dagger \delta_q v &=& i\Delta^\dagger_-\delta_q\Delta_-+i\Delta^\dagger_+\delta_q\Delta_+\\
&&+i\int^{-\lambda}_{-\infty} ds \psi^\dagger_L(s)\delta_q\Psi_L(s)+\int^\lambda_{-\lambda} ds \psi^\dagger_M(s)\delta_q\psi_M(s)+\int^\infty_\lambda ds \psi^\dagger_R(s)\delta_q\psi_R(s) \nonumber
\end{eqnarray}
which after substitution gives us
\begin{eqnarray}
iv^\dagger \delta_q  v &=& \sqrt{\frac{\mathcal{L}}{\mathcal{D}r}}\nonumber\\
&&+(i\int_{-\infty}^{-\lambda}ds \psi_L^\dagger\psi_L+i\Delta^\dagger_+\Delta_-)(e^{\vec{\sigma}\cdot\vec{r}\lambda}\delta_qe^{-\vec{-\sigma}\cdot\vec{r}\lambda}-\frac{1}{\mathcal{D}} \zeta^\dagger_-\delta_qf_-f^\dagger_-\zeta_-)\nonumber\\
&&+(i\int_{\lambda}^{\infty}ds \psi_R^\dagger\psi_R+i\Delta^\dagger_+\Delta_+)(e^{-\vec{\sigma}\cdot\vec{r}\lambda}\delta_qe^{-\vec{-\sigma}\cdot\vec{r}\lambda}-\frac{1}{\mathcal{D}} \zeta^\dagger_+\delta_qf_+f^\dagger_+\zeta_+)\nonumber\\
&&+\frac{r}{\mathcal{L}}(e^{-\vec{\sigma}\cdot\vec{r}\lambda}f_+\delta_qf^\dagger_+e^{-\vec{\sigma}\cdot\vec{r}\lambda}+e^{\vec{\sigma}\cdot\vec{r}\lambda}f_-\delta_qf^\dagger_-e^{\vec{\sigma}\cdot\vec{r}\lambda})\nonumber\\
&&+\int ds \psi^\dagger_M\psi_M e^{-\vec{\sigma}\cdot\vec{r}s}\delta_qe^{\vec{\sigma}\cdot\vec{r}s}
\end{eqnarray}

We compute the following
\begin{eqnarray}
i\Delta^\dagger_-\Delta_-+i\int ds \psi_L^\dagger \psi_L &=& i\frac{r}{\mathcal{L}}(M-N\vec{\sigma}\cdot\hat{r})\nonumber\\
i\Delta^\dagger_+\Delta_++i\int ds \psi_R^\dagger\psi_R &=& i\frac{r}{\mathcal{L}}(M+N\vec{\sigma}\cdot\hat{r})\nonumber\\
i\frac{r}{\mathcal{L}} \left(e^{-\vec{\sigma}\cdot\vec{r}\lambda}f_+\delta_qf^\dagger_+e^{-\vec{\sigma}\cdot\vec{r}\lambda}+e^{\vec{\sigma}\cdot\vec{r}\lambda}f_-\delta_qf^\dagger_-e^{\vec{\sigma}\cdot\vec{r}\lambda} \right) &=& (i C \hat{d}_q-i S \hat{r}_q - C\vec{\sigma}\cdot\vec{R}^q\nonumber\\
&&+(2S\omega_q d-C\hat{r}\cdot\vec{R}^q d+\hat{r}\cdot\vec{R}^q d)\vec{\sigma}\cdot\hat{r})\nonumber
\end{eqnarray}

\begin{eqnarray}
(i\Delta^\dagger_-\Delta_-+i\int ds \psi^\dagger_L\Psi_L) e^{\vec{\sigma}\cdot\vec{r}\lambda}\delta_qe^{-\vec{\sigma}\cdot\vec{r}\lambda}+(i\Delta^\dagger_+\Delta_++i\int ds \psi^\dagger_R\psi_R)e^{-\vec{\sigma}\cdot\vec{r}\lambda}\delta_qe^{\vec{\sigma}\cdot\vec{r}\lambda}
\nonumber\\
= \frac{i}{\mathcal{L}}\left(-N(2\lambda r_q)+(MC-M-SN)i(\vec{\sigma}\times\hat{r})_q\right)\nonumber
\end{eqnarray}

\begin{equation}
i\int ds \psi^\dagger_M\psi_M e^{-\vec{\sigma}\cdot\vec{r}s}\delta e^{\vec{\sigma}\cdot\vec{r}s} =-i\frac{\mathcal{D}}{2\mathcal{L}}\left(2\hat{r}_q(\lambda C-\frac{S}{2r})+\frac{1}{r}(S-2\lambda r)i(\vec{\sigma}\times\hat{r})_q\right)\nonumber
\end{equation}

\begin{eqnarray}
\sqrt{\frac{\mathcal{L}}{\mathcal{D}r}}\delta_q\sqrt{\frac{\mathcal{D}r}{\mathcal{L}}}&=&\frac{2(a\hat{d}_q-r_q)}{\mathcal{D}}-\frac{\hat{r}_q}{r}\nonumber\\
&&+2\lambda \hat{r}_q \frac{\mathcal{K}}{\mathcal{L}} -\frac{2}{\mathcal{L}}\left( (r\hat{d}_q-a\hat{r}_q)C+(r_q+a\hat{d}_q) S\right) \nonumber
\end{eqnarray}

\begin{eqnarray}
-\frac{1}{\mathcal{D}}(\zeta^\dagger_-\delta_qf_-f_-^\dagger\zeta_-)(i\Delta^\dagger_-\Delta_-+i\int ds\psi_L^\dagger\psi_L)-\frac{1}{\mathcal{D}}(\zeta_+^\dagger\delta_qf_+f^\dagger_+\zeta_+)(i\Delta_+^\dagger\Delta_++i\int ds \psi_R^\dagger \psi_R)\nonumber\\
=-i\frac{r}{\mathcal{D}\mathcal{L}}\left( 2M(z\hat{d}_q+z_q)+N(-i4z\omega_q d+i2\vec{z}\cdot\vec{R}^q d)\vec{\sigma}\cdot\hat{r}\right)\nonumber
\end{eqnarray}

where we have used the following results
\begin{eqnarray}
f_+\delta_qf^\dagger_++f_-\delta_qf^\dagger_-=\hat{d}_q+i\vec{\sigma}\cdot\vec{R}_qd\nonumber\\
f_+\delta_q f^\dagger_++\delta_q f_+f^\dagger_+=\hat{d}_q+\sigma_q\nonumber\\
f_+\delta_q f^\dagger_+-f_-\delta f^\dagger_- = -i R^q_r d+\sigma_q = i2\omega_q d+\sigma_q\nonumber\\
f_+ \delta_q f^\dagger_+-\delta_q f_+f^\dagger_+=i2\omega_q d-i\vec{\sigma}\cdot\vec{R}^qd
\end{eqnarray}
where $\vec{R}_q$ are the canonical right one-forms on the group $SU(2)$, also known as the Maurer-Cartan one-forms. $R^q_3=-2\omega_q$ where $\vec{\nabla}\times\vec{\omega}=\nabla (\lambda+\frac{1}{2d})$.

As a reminder we have $a=z+d$, $\mathcal{D}=a^2-r^2$, $C=\cosh(2\lambda r)$, $S=\sinh(2\lambda r)$, $M=aC+rS$, $N=aS+rC$, $\mathcal{K}=(a^2+r^2)C+2raS$, and $\mathcal{L}=(a^2+r^2)S+2raC$.

We combine these intermediate terms to construct $iv^\dagger \delta_q v$.
\begin{eqnarray}
i v^\dagger \delta_q v &=& \frac{r}{\mathcal{L}} \{\vec{\sigma}\cdot\hat{r} (-2 \sinh(2\lambda r) \omega_q d-\cosh(2\lambda r) \hat{r}\cdot\vec{R}_q d \nonumber\\
&&+\hat{r}\cdot\vec{R}_q d - \frac{N}{\mathcal{D}}\left( 4z\omega_q d-2\vec{z}\cdot\vec{R}_q d\right) ) \nonumber\\
&&-(\vec{\sigma}\times\hat{r})_q\left(\frac{a}{r} - \frac{a}{r} \cosh(2\lambda r) - \sinh(2\lambda r)+\frac{\mathcal{D}}{r}\left(\lambda-\frac{\sinh(2\lambda r)}{2r}\right)\right)\nonumber\\
&& -\vec{\sigma}\cdot\vec{R}_q d\}
\end{eqnarray}


\addtocontents{toc}{\vspace{2em}}  
\backmatter

\label{Bibliography}
\lhead{\emph{Bibliography}}  
\bibliographystyle{unsrtnat}  

\end{document}